\documentclass[preprint2,times,tighten]{aastex6}

\usepackage[all]{hypcap} 

\usepackage{amsmath,amsfonts, amssymb}

\usepackage{mathrsfs}

\usepackage{dsfont}

\usepackage{url}
\usepackage{paralist}

\usepackage{float}
\restylefloat{deluxtable}

\usepackage{subfigure}

\shorttitle{The Geometry of Sagittarius Stream from Pan-STARRS1 3$\pi$ RR Lyrae}
\shortauthors{Hernitschek et al.}

\begin{document}


\title{The Geometry of Sagittarius Stream from Pan-STARRS1 3$\pi$ RR Lyrae}


\author{Nina Hernitschek\altaffilmark{1,2}, Branimir Sesar\altaffilmark{2}, Hans-Walter Rix\altaffilmark{2}, Vasily Belokurov\altaffilmark{3}, 
David Martinez-Delgado\altaffilmark{4}, Nicolas F. Martin \altaffilmark{5,2}, Nick Kaiser\altaffilmark{6}, Klaus Hodapp\altaffilmark{6}, Kenneth C. Chambers\altaffilmark{6}, Richard Wainscoat \altaffilmark{6}, Eugene Magnier\altaffilmark{6}, Rolf-Peter Kudritzki\altaffilmark{6}, Nigel Metcalfe\altaffilmark{7}, Peter W. Draper\altaffilmark{7}}
\email{ninah@astro.caltech.edu}

\altaffiltext{1}{Division of Physics, Mathematics and Astronomy, Caltech, Pasadena, CA 91125}    
\altaffiltext{2}{Max-Planck-Institut f{\"u}r Astronomie, K{\"o}nigstuhl 17, 69117 Heidelberg, Germany}    
\altaffiltext{3}{Institute of Astronomy, University of Cambridge, Madingley Road, Cambridge CB3 0HA, UK}
\altaffiltext{4}{Astronomisches  Rechen-Institut,  Zentrum  f\"{u}r  Astronomie der Universit\"{a}t Heidelberg,  M\"{o}nchhofstr. 12-14,  69120 Heidelberg, Germany}
\altaffiltext{5}{Observatoire astronomique de Strasbourg, Universit\'{e} de Strasbourg, CNRS, UMR 7550, 11 rue de l'Universit\'{e}, F-6700 Strasbourg, France}
\altaffiltext{6}{Institute for Astronomy, University of Hawai’i at Manoa, Honolulu, HI 96822, USA}
\altaffiltext{7}{Department of Physics, University of Durham, South Road, Durham DH1 3LE, UK}



\begin{abstract}
We present a comprehensive and precise description of the Sagittarius (Sgr) stellar stream's 3D geometry
as traced by its old stellar population. This analysis draws on the sample of ${\sim}44,000$ RR Lyrae (RRab) stars from the Pan-STARRS1 (PS1) 3$\pi$ survey \citep{Hernitschek2016,Sesar2017b}, which is ${\sim}80\%$ complete and ${\sim}90\%$ pure within 80~kpc, and extends to ${\gtrsim} 120$~kpc with a 
distance precision of ${\sim} 3\%$. A projection of RR Lyrae stars within $|\tilde{B}|_{\odot}<9^\circ$ of the Sgr stream's orbital plane reveals the morphology of both the leading and the trailing arms at very high contrast, across much of the sky. In particular, the map traces the stream near-contiguously through the distant apocenters. We fit a simple model for the mean distance and line-of-sight depth of the Sgr stream as a function of the orbital plane angle $\tilde{\Lambda}_{\odot}$, along with a power-law background-model for the field stars. This modeling results in estimates of the mean stream distance precise to ${\sim}1\%$ and it resolves the stream's line-of-sight depth. These improved geometric constraints can serve as new constraints for dynamical stream models.
\end{abstract}

\section{Introduction}
\label{sec:Introduction}

Stellar streams around galaxies, and in particular around the Milky Way, are of 
great interest as their orbits are sensitive tracers of a galaxy's formation 
history and gravitational potential \citep[e.g.][]{Eyre2009, Law2010, Newberg2010, Sanders2013}. In the Milky Way, the 
Sagittarius (Sgr) stream is the dominant tidal stellar stream
of the Galactic stellar halo,
and its extent has been traced around much of the sky. The stream shows two pronounced tidal tails extending each ${\sim}180 \arcdeg$ and reaching Galactocentric distances from 20 to more than 100~kpc, also referred to as ``leading" and ``trailing arm" \citep{Majewski2003}. 

Stellar streams are sets of stars on similar orbits and therefore lend 
themselves to constraining the dynamical mass within their orbit. 
The distribution of Sgr stream's stars can therefore serve as a probe of 
the Galactic mass profile and shape, including the dark matter halo.
This is best done with 6-dimensional phase-space information available for the stars, 
as has been shown for relatively nearby streams  such 
as GD-1 \citep{Koposov2010, Bovy2016} and Ophiuchus \citep{Sesar2016}.

Since its discovery by \cite{Ibata1994}, several work on sections of the Sagittarius stream was carried out. 
The first modeling attempt was done by \cite{Johnston1995}, however, finding the progenitor, the Sagittarius dwarf galaxy, disrupting after only two orbits while observations show the completion of about 10 orbits. As a solution to the problem, \cite{Ibata1998} concluded from an extensive numerical study that the Sagittarius dwarf galaxy must have a stiff and extended dark matter halo if it has still about 25 \% of its initial mass and is still bound today.

Early pencil-beam surveys before the large-scale survey era were used by \cite{Mateo1998}, \cite{Martinez2001} and \cite{Martinez2004}, reporting detections of tidal debris in the northern stream of the Sagittarius dwarf galaxy and leading to the publication of one of the first models of the Sagittarius stream being in good agreement with the observations \citep{Martinez2004}.
Since the first detailed mapping by \cite{Majewski2003}, there have been quite a number of attempts to map and trace the Sgr stream over larger fractions of its extent, at least in part e.g. building on the seminal work by \cite{Majewski2003}.
Such work was carried out by \cite{NiedersteOstholt2010}, who traced the Sgr stream out to $D{\sim}50$~kpc using main sequence, red giant, and horizontal branch stars from the SDSS as well as M giants from the Two Micron All-Sky Survey (2MASS), \cite{Koposov2012} who used main-sequence turn-off (MSTO) stars to measure the stream's
distance gradients between $\tilde{\Lambda}_{\odot} = 90\arcdeg - 130\arcdeg$ in the sourthern Galactic hemisphere, and \cite{Slater2013} using color-selected MSTO stars from the Pan-STARRS1 surved to present a panoramic view of the Sgr tidal stream in the southern Galactic hemisphere spanning $\tilde{\Lambda}_{\odot} = 70\arcdeg - 130\arcdeg$.

Wide area surveys of the Galactic halo, employing RR Lyrae as tracers, have already been used in the past: \cite{Vivas2001} carried out a study on 148 RR Lyrae within the first 100 deg$^2$ of the Quasar Equatorial Survey Team (QUEST) RR Lyrae survey, and after publishing a catalog \citep{Vivas2004} continued using QUEST for finding substructure near the Virgo overdensity \citep{Vivas2008}. \cite{Duffau2014} (with Vivas) have extended the sample and found various velocity groups from QUEST and QUEST-La Silla \citep{Zinn2014}. \cite{Sesar2012} found two new halo velocity groups using RR Lyrae from the Palomar Transient Factory (PTF) survey, \cite{Sesar2013b} used a sample of ${\sim}$5000 RR Lyrae over ${\sim}$8000 deg$^2$ of sky from the the Lincoln Near-Earth Asteroid Research asteroid survey (LINEAR) survey to analyze the Galactic stellar halo profile for heliocentric distances between 5 kpc and 30 kpc. \cite{Drake2014} produced a catalog of RR Lyrae and other periodic variables from the Catalina Surveys Data Release-1 (CSDR1).

A number of these attempts have been able to map parts of the Sagittarius stream. An extensive map was made by \citep{Drake2013a, Drake2014}, which confirms the presence of a halo structure that appears as part of the Sagittarius tidal stream, but is inconsistent with N-body simulations of that stream like the \cite{Law2010} model. Shortly before, this feature was confirmed by \cite{Belokurov2014} based on M-giants.

In more recent work, \cite{Belokurov2014} have demonstrated that the trailing arm of the Sgr stream can be traced out to its apocenter at ${\sim}100$~kpc. They also give a fit of the stream's leading arm to its apocenter at ${\sim}50$~kpc. The extent of the Sgr stream has therefore only recently became fully apparent, spanning an unparalleled range of distances when compared to other stellar tidal streams in the Milky Way.

In contrast to the aforementioned partial mapping of the Sgr stream, showing the stream only piecewise mapped by tracers from different surveys and often relying on different kinds of sources as tracers, the data we have at hand -- RR Lyrae stars from Pan-STARRS1 -- enables us to trace the complete angular extent of the Sgr stream as well as to look even to the outskirts of the stream.

There have also been attempts to model the 
Sagittarius tidal stream \citep[e.g.][]{Law2005b, Penarrubia2010, Gibbons2014}, 
which has complex geometry and incomplete (so far) phase-space information.

\cite{Helmi2004a} and \cite{Helmi2004b} claim that the trailing arm is too young to be a probe of the dark matter profile, whileas the leading arm, being slightly older, provides a direct evidence for the prolate shape of the dark matter halo.
\cite{Helmi2004a} and \cite{Helmi2004b} have used numerical simulations of the Sgr stream to probe the profile of the Milky Way's dark matter halo.
They find that the data available for the stream are consistent with a Galactic dark matter halo that could be either oblate or prolate, with minor-to-major density
axis ratios can be as low as 0.6 within the region probed by the Sgr stream.
In agreement with \cite{Martinez2004}, they state that the dark matter halo should thus not be assumed as nearly spherical.

The modelling efforts have also included $N$-body simulations constrained by observational data \citep[e.g.][]{Fellhauer2006, Law2010, Penarrubia2010, Dierickx2017}.
Consistent 3D stream constraints from a single survey, as we set out to do here, aids the comparison to models of the Sgr stream, usually based on N-body simulation \citep[e.g.][]{Law2010,Dierickx2017}.

The main aim of this paper is to map the geometry (in particular the distance) of 
the Sagittarius (Sgr) stream more precisely, accurately and comprehensively than 
before, using exclusively RR Lyrae stars (RRL) from a single survey to trace the stream's old stellar population.
For our analysis, we use the RRab sample of \cite{Sesar2017b}, which covers $3/4$ of the sky, is rather pure and has precise distances (to $3\%$).
It was generated from data of the Pan-STARRS1 survey (PS1) \citep{Kaiser2010}, using structure functions and a machine-learning algorithm by \cite{Hernitschek2016} and a subsequent multi-band light-curve fitting and another machine-learning algorithm as described in \cite{Sesar2017b}. 

This provides us with an RRL map of the old Galactic stellar halo
that is of high enough contrast to fit the Sgr stream geometry directly by a density model: its distance and line-of-sigh-depth as a function 
of angle in its orbital plane. In particular, we can derive precise apocenter positions of both the leading and trailing arms and thus the Galactocentric orbital precession of the stream.

The structure of the paper is as follows: in Section~\ref{sec:PS1} we describe 
the PS1 survey and the RRL sample derived from it; in Section~\ref{sec:TheModelSgr}
we describe and apply the distance distribution model for the Sgr stream we fit to these 
data; in Section~\ref{sec:results} we present and discuss our results obtained from evaluating the fit, describe our findings regarding geometrical properties of the stream
and compare them to earlier work; we conclude with a discussion and summary in Section \ref{sec:discussion}.\\

This work is part of a series of papers exploring the identification 
and 
astrophysical exploitation of RRL stars in the PS1 survey. The basic approach for 
applying multi-band structure functions to PS1 3$\pi$ lightcurves, and 
subsequently using a classifier evaluating variability and color information to 
select RR Lyrae and QSO candidates has been laid out in \cite{Hernitschek2016}, 
with results from the preliminary PS1 3$\pi$ version, PV2. We then applied
multi-band period fitting to all these RRL candidates \citep{Sesar2017b}, using light curves from the 
final PS1 3$\pi$ version, PV3. The quality and plausibility of these fits aided in the
classification, increasing the purity of the sample and leading to 
precise distance estimates for the sample of RRab stars.
\cite{Sesar2017c} shows new detections within the Sgr stream, made using the RRab sample without further fitting or modeling; in particular, they show the detection of spatially distinct ``spur'' and clump features reaching out to more than 100 kpc on top of the apocenters of the Sgr stream, being in good agreement with recent dynamical models \citep{Gibbons2014,Fardal2015,Dierickx2017}.

\section{RR Lyrae Stars from the PS1 Survey}
\label{sec:PS1}

Our analysis is based on a sample of highly likely RRab stars, as selected by \cite{Sesar2017b} from the Pan-STARRS1 3$\pi$ survey. 
In this section, we describe the pertinent properties of the PS1 3$\pi$ survey and its obtained light curves, recapitulate briefly the process of selecting the likely RRab, as laid out in \cite{Sesar2017b}; and we  briefly characterize the obtained candidate sample.

The Pan-STARRS 1 (PS1) survey \citep{Kaiser2010} is collecting multi-epoch, multi-color observations undertaking a number of surveys, among which the PS1 3$\pi$ survey \citep{Stubbs2010, Tonry2012,Chambers2016} is currently the largest. It has observed the entire sky north of declination $-30\arcdeg$ in five filter bands ($g_{\rm P1},r_{\rm P1},i_{\rm P1},z_{\rm P1},y_{\rm P1}$) 
with a 5$\sigma$ single epoch depth of 22.0, 21.8, 21.5, 20.9 and 19.7 magnitudes in $g_{\rm P1},r_{\rm P1},i_{\rm P1},z_{\rm P1}$, and $y_{\rm P1}$, respectively  \citep{Chambers2016}.

For more than $1.1\times 10^9$ PS1 3$\pi$ PV3 sources, we constructed a set of data features for source classification: the sources' mean magnitudes in various bands, as well as multi-band variability features like a simple $\chi^2$-related variability measure $\hat{\chi}^2$,  
and multi-band structure function parameters, $(\omega_r,\tau)$, describing the characteristic variability amplitude and time-scale \citep{Hernitschek2016}. Based on these features, including a multi-band light-curve fit resulting in period estimates, a machine-learned classifier, trained on PS1 3$\pi$ sources within SDSS S82, then selects plausible RRL candidates \citep{Sesar2017b}. Their distances were calculated based on a newly derived period-luminosity relation for the optical/near-infrared PS1 bands, as the majority of the PS1 sources lack metallicities. The complete methodology on how to derive the distances and verify their precision is given in \cite{Sesar2017b}.

Overall, this highly effective identification of RR Lyrae stars has resulted in the widest (3/4 of the sky) and deepest (reaching $> 120$~kpc) sample of those stars to date. 
The RRab sample from \cite{Sesar2017b} were selected uniformly from the set of sources in the PS1 3$\pi$ survey in the 
area and apparent magnitude range available for this survey. \cite{Sesar2017b} have shown that the selection completeness and purity 
for sources at high galactic latitudes ($\vert b \vert > 15\arcdeg$) is approximately uniformly over a wide range 
of apparent magnitude up to a flux-averaged r-band magnitude of 20 mag,
maintaining a sample completeness for the RRab stars of $\sim $80\% and a purity of $\sim $90\% within 80~kpc (see their Fig. 11).

We thus explicitly refer to high-latitude completeness on PS1 3$\pi$ overlapping with SDSS Stripe 82 \citep{Sesar2017b}, but we have no reason to believe that the purity and completeness varies strongly across high-latitude areas. A detailed map of the purity and completeness including not only their distance but spatial distribution would require that we have ``ground truth" (i.e.: knowledge about the true type of star for every source) in all directions, which is of course not available.
For the definition of completeness and purity, we refer to \cite{Sesar2017b}, where the completeness is defined as the fraction of recovered RR Lyrae stars on a test area (e.g. SDSS Stripe 82), and the purity is defined as the fraction of true RR Lyrae
stars in the selected sample of RR Lyrae candidates.

There are 44,403 likely RRab stars in this PS1 sample with distance estimates that are precise to $3\%$. In the further analysis, we refer to this sample 
\citep{Hernitschek2016,Sesar2017b} as ``RRab stars''.

\begin{figure*}
\begin{center}  
\includegraphics[]{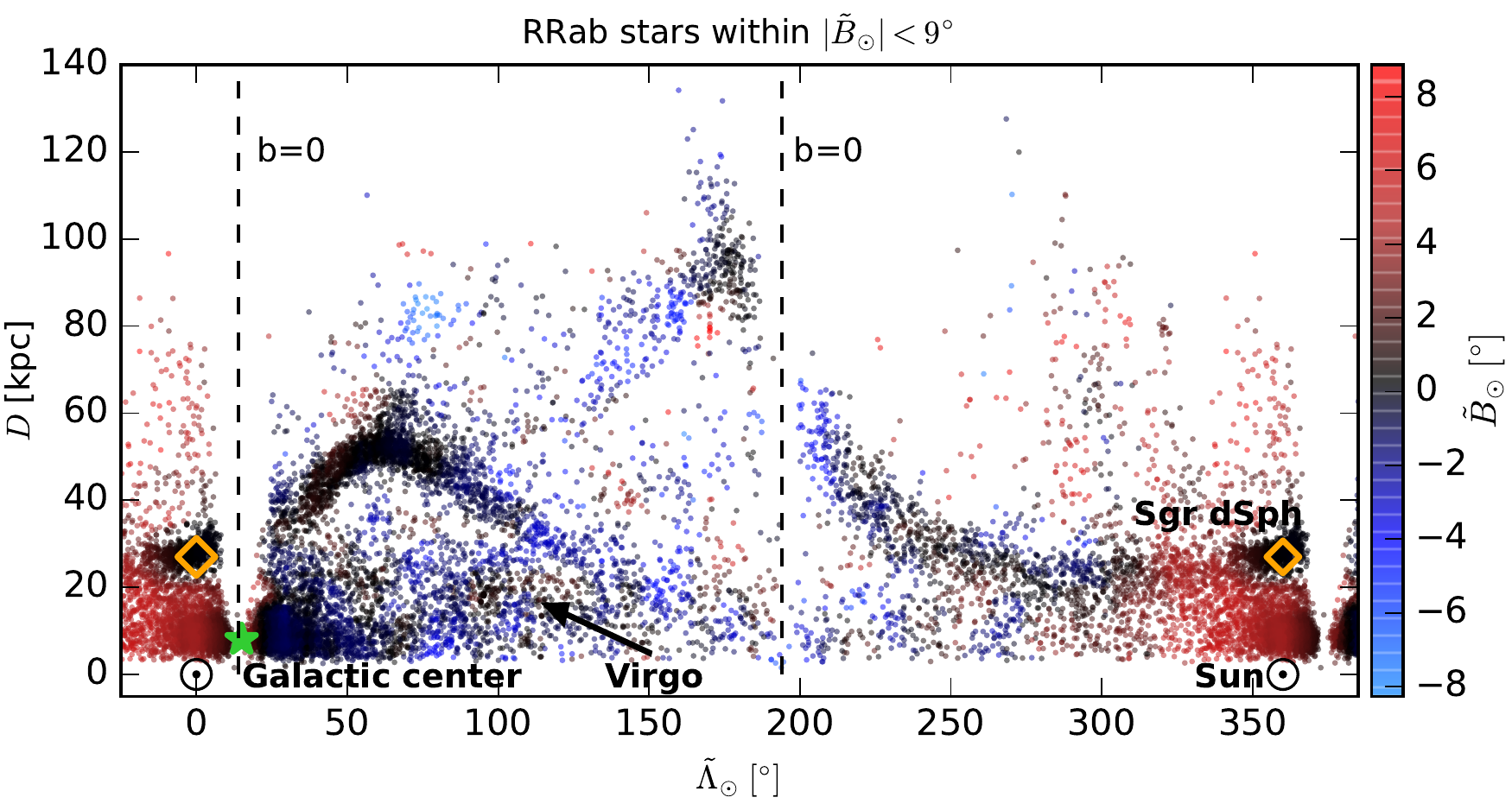}
\caption[RRab stars within $\vert{\tilde{B}_{\odot}}\vert<9\arcdeg$ as obtained after period fitting \citep{Sesar2017b}.]
{{RRab stars within $\vert{\tilde{B}_{\odot}}\vert<9\arcdeg$ as obtained after period fitting \citep{Sesar2017b}. The Sgr stream is clearly visible up to ${\sim}130$~kpc. The color indicates the median angular distance $\tilde{B}_{\odot}$ of a $5\arcdeg\times5$ kpc bin 
(in $\tilde{\Lambda}_{\odot}$ and $D$ coordinates) from the Sgr orbital plane $\tilde{B}_{\odot}=0\arcdeg$. This was chosen due to the high source density in some regions.\newline
In this figure, the angular coordinate $\tilde{\Lambda}_{\odot}$ is running from $-20\arcdeg$ to $380\arcdeg$ with repeated data points for $\tilde{\Lambda}_{\odot}<0\arcdeg$ and $\tilde{\Lambda}_{\odot}>360\arcdeg$, to better show the distribution near $\tilde{\Lambda}_{\odot} \sim 0\arcdeg$.\newline
The location of the Sun, Galactic anticenter, Sgr dSph and the Virgo overdensity \citep{Vivas2001, Newberg2002, Juric2008} are indicated. The dashed line marks the position of the Galactic plane. The centroid for Sgr dSph was taken from \cite{Karachentsev2004}.\newline
The Cetus stream should cross the Sgr stream at $\tilde{\Lambda}_{\odot} \sim 270 \arcdeg$, ${\tilde{B}_{\odot}} \sim 1 \arcdeg$ \citep{Newberg2009}. From our data, every evidence is marginal.
}
\label{fig:sgr_plot_candidates_sesar_for_paper}}
\end{center}
\end{figure*}

While the sample covers the entire sky above 
$\mathrm{Dec} > -30^\circ$, we focus on stars near the Sgr stream orbital plane. 
We use the heliocentric Sagittarius coordinates $(\tilde{\Lambda}_{\odot}, \tilde{B}_{\odot})$ as defined by \cite{Belokurov2014}, where the equator $\tilde{B}_{\odot}=0\arcdeg$ is aligned with the plane of the stream.  We restrict our subsequent analysis to RRab from our sample that lie within $\vert \tilde{B}_{\odot} \vert <9\arcdeg$ as also seen in the plots by \cite{Belokurov2014}, resulting into ${\sim}15,000$ stars. 
This sample is plotted in Fig. \ref{fig:sgr_plot_candidates_sesar_for_paper}
in the $(\tilde{\Lambda}_{\odot},D)$ plane of longitudinal coordinates $\tilde{\Lambda}_{\odot}$ and heliocentric distances $D$, with the angular distance to the Sgr plane $\tilde{B}_{\odot}$ indicated by color-coding.
A table for these stars within $\vert \tilde{B}_{\odot} \vert <-9\arcdeg$ is given in the Appendix, Tab. \ref{tab:table_RRLyr}. A machine readable version of this table is available in the electronic edition of the Journal.

\section{A Simple Model to Characterize the Sgr Stream Geometry}
\label{sec:TheModelSgr}

We aim at a simple quantitative description of the Sgr stream, by providing the mean distance and line-of-sight (l.o.s.) depth of presumed member stars, as a function of angle in the orbital plane. We only consider stars within $|\tilde{B}_{\odot}|<9\arcdeg$, and marginalize over their distribution perpendicular to the orbital plane, resulting in 
a set of distances as a function of $\tilde{\Lambda}_{\odot}$. 
In practice, the overall distance distribution $p_{\mathrm{RRL}}(D)$ towards any $\tilde{\Lambda}_{\odot}$ is modeled as the superposition of a ``stream'' and ``halo'' component. At each $\tilde{\Lambda}_{\odot}$ bin, the halo is modeled as a power-law $\rho_{\mathrm{halo}}$ in Galactic coordinates, describing the background of field stars.
The heliocentric distance distribution of stream stars is modeled as a Gaussian, characterized by 
$D_{\mathrm{sgr}}$ and the l.o.s. depth, $\sigma_{\mathrm{sgr}}$:
\begin{align}
p_{\mathrm{RRL}}(\mathcal{D}|\boldsymbol{\theta}) &=& p_{\mathrm{halo}}(\mathcal{D}|\boldsymbol{\theta})+p_{\mathrm{stream}}(\mathcal{D}|\boldsymbol{\theta})\label{eq:stream_halo_model}\\
&=&(1-f_{\mathrm{sgr}})\times \hat{\rho}_{\mathrm{halo}}(l,b,D,q,n) \nonumber \\
& &+ f_{\mathrm{sgr}} \times \hat{\rho}_{\mathrm{sgr}}(l,b,D,D_{\mathrm{sgr}},\sigma_{\mathrm{sgr}}),\nonumber
\end{align}
where 
\begin{equation}
\hat{\rho}_{\mathrm{halo}}(l,b,D,q,n)\equiv \frac{\rho_{\mathrm{halo}}(l,b,D,q,n)}{\int\limits_{D_{min}}^{D_{max}}\rho_{\mathrm{halo}}(l,b,D,q,n)\mathrm{d}D},
\end{equation}
with an analogous definition of $\hat{\rho}_{\mathrm{sgr}}$.
The data set is given as $\mathcal{D} = (D, \delta D, l,b)$. The parameters are $\boldsymbol{\theta}=(f_{\mathrm{sgr}},D_{\mathrm{sgr}},\sigma_{\mathrm{sgr}},n)$, composed of the fraction of the stars $f_{\mathrm{sgr}}$ being in the Sgr stream at the given $\tilde{\Lambda}_{\odot}$ slice, the heliocentric distance of the stream $D_{\mathrm{sgr}}$, its l.o.s. depth $\sigma_{\mathrm{sgr}}$, and the power-law index $n$ of the halo model. $D_{\mathrm{min}}$ and $D_{\mathrm{max}}$ are the minimum and maximum $D$ 
we consider in each $\tilde{\Lambda}_{\odot}$ slice.
	
We adopt a simple power-law halo model $\rho_{\mathrm{halo}}$ \citep{Sesar2013b} to describe the ``background" of field stars in the direction of $(l,b)$:
\begin{equation}
\rho_{\mathrm{halo}}(X,Y,Z)=\rho_{\odot \mathrm{RRL}}\left(R_{\odot}/r_q \right)^n
\label{eqn:halomodel}
\end{equation}
with
\begin{align*}
X &= R_{\odot} - D \cos l \cos b \\
Y &= - D \sin l \cos b \\
Z &= D \sin b \\
r_q &= \sqrt{X^2 + Y^2 + (Z/q)^2}. 
\end{align*}

\cite{Sesar2013b} also give the halo parameters
\begin{align*}
n &= 2.62 \\
R_{\odot} &= 8.0~\mathrm{kpc} \\
q &= 0.71 \\
\rho_{\odot \mathrm{ RRL}} &= 4.5~\mathrm{kpc}^{-3}.
\end{align*}

Here, $\rho_{\odot \mathrm{RRL}}$ is the number density of RR Lyrae at the position of the Sun, $q$ gives the halo flattening along the $Z$ direction.
In our analysis, all ``background halo" parameters except the fitting parameter $n$ are kept fixed.

The stream is modeled as a normal distribution centered on $D_{\mathrm{sgr}}$ and with variance $\sigma_{\mathrm{sgr}}$ as follows. It is defined in Galactic coordinates $(l,b)$ and Galactocentric distance $R$, where $R$ is given as function of the heliocentric distances $D$, $D_{\mathrm{sgr}}$, and distance uncertainty $\delta D$, as follows:
\begin{align}
& \rho_{\mathrm{sgr}}(l,b,D,\delta D,D_{\mathrm{sgr}},\sigma_{\mathrm{sgr}})\nonumber \\ & =\frac{1}{\sqrt{2\pi(\sigma_{\mathrm{sgr}}^2+\delta D^2)}} \exp \left(  -\frac{ (R(D)-R(D_{\mathrm{sgr}}))^2}{2(\sigma_{\mathrm{sgr}}^2+ \delta D^2)}\right) D^2.
\label{eqn:streammodel}
\end{align}

For the distance uncertainties of RRab stars, we adopt a $\delta D$ of $3\%$ according to \cite{Sesar2017b}. 

\subsection{Fitting the Sgr Model}
\label{sec:Fitting}
For fitting this model, the sample of RRab stars near the Sgr orbital plane is split into bins of $\tilde{\Lambda}_{\odot} \pm \frac{\Delta \tilde{\Lambda}_{\odot}}{2}$, each $\Delta \tilde{\Lambda}_{\odot} =10\arcdeg$ wide; the data are not binned in $D$. In each bin, we fit (independently) the parameters of the stream, $D_{\mathrm{sgr}}$ and $\sigma_{\mathrm{sgr}}$, along with the halo model parameter $n$. Whereas it is obvious why the stream-related model parameters should be fitted individually for each $\tilde{\Lambda}_{\odot}$ bin, the reason for fitting also the halo power law index $n$ individually is to account for incompleteness of the data. The flattening parameter $q$ is kept fixed at 0.71, as fitting for $q$ did not improve the results for the stream-related model parameters.

To constrain the geometry of the Sgr stream in a probabilistic manner, we 
calculate the joint posterior probability $p_{\mathrm{RRL}}(\boldsymbol{\theta} \vert \mathcal{D})$ of the parameter set 
$\boldsymbol{\theta}=(f_{\mathrm{sgr}},D_{\mathrm{sgr}},\sigma_{\mathrm{sgr}},n)$, given the data set $\mathcal{D} = (D, \delta D, l,b)$. The marginal posterior probability of the parameter set $\boldsymbol{\theta}$, $p_{\mathrm{RRL}}(\boldsymbol{\theta} \vert \mathcal{D})$ is related to the marginal likelihood $p_{\mathrm{RRL}}(\mathcal{D} \vert \boldsymbol{\theta})$ through 
\begin{equation}
p_{\mathrm{RRL}}(\boldsymbol{\theta} \vert \mathcal{D}) \propto p_{\mathrm{RRL}}(\mathcal{D} \vert \boldsymbol{\theta}) p(\boldsymbol{\theta})
\label{equ:posteriorprobability}
\end{equation}   
where $p(\boldsymbol{\theta})$ is the prior probability of the parameter value set.

We evaluate 
\begin{align}
\ln p_{\mathrm{RRL}}(\boldsymbol{\theta} \vert \mathcal{D}) & = \sum_{i} \ln p_{\mathrm{RRL}}(\mathcal{D}_i \vert \boldsymbol{\theta}) + \ln p (\boldsymbol{\theta})
\label{equ:evaluate}
\end{align}   
with $p_{\mathrm{RRL}}(\mathcal{D}_i|\boldsymbol{\theta})$ given by Equ. \eqref{eq:stream_halo_model}, and $i$ indexes the RRab stars.

We use the following prior probability for the model parameters, $p(\boldsymbol{\theta})$:
for $\sigma_{\mathrm{sgr}}$, we choose a prior that is uniform in $\ln$, whereas for the other parameters, we adopt uniform priors.
Specifically, we adopt

\begin{align}
\ln p(\boldsymbol{\theta}) =& -\ln(\sigma_{\mathrm{sgr}}) \\& + \mathrm{Uniform}( 0.05 \leq f_{\mathrm{sgr}} < 1) \nonumber \\&+ \mathrm{Uniform}( 1.7  \leq n < 5.0) \nonumber \\&+ p(D_{\mathrm{sgr}} \vert \tilde{\Lambda}_{\odot}) \nonumber. 
\label{eqn:prior_1}
\end{align}

The prior for $D_{\mathrm{sgr}}$ depends on $\tilde{\Lambda}_{\odot}$, and is uniform within $D_{\mathrm{minprior}}(\tilde{\Lambda}_{\odot})$, $D_{\mathrm{maxprior}}(\tilde{\Lambda}_{\odot})$ as indicated in Fig. \ref{fig:bestfit} and listed in Tables \ref{tab:Dprior_leading} and \ref{tab:Dprior_trailing}.
Whereas the prior is generally wide, a quite restrictive prior was chosen for $20\arcdeg \leq \tilde{\Lambda}_{\odot}<30 \arcdeg$
and $30\arcdeg \leq \tilde{\Lambda}_{\odot}<40 \arcdeg$, as the fit otherwise behaves poorly because of the background sources along these lines of sight.

$D_{\mathrm{minprior}}$, $D_{\mathrm{maxprior}}$ are basically constrained by the minimum and maximum distance in the $\tilde{\Lambda}_{\odot}$ slice in case, but are also defined in order to mask dense regions at low heliocentric distances as well as to separate the leading and trailing arm where both are present at the same line of sight.

The most probable model given the data is explored using the Affine Invariant Markov Chain Monte Carlo (MCMC) ensemble sampler \citep{Goodman2010} as implemented in the \texttt{emcee} package \citep{Foreman2012}.

The approach was verified with mock data, using a halo component that was sampled from the underlying halo model, 
superimposed by a mock stream that was inserted as a stellar density sheet; its number density is uniform perpendicular to the line of sight, and Gaussian along the line of sight.
The fraction of the stream stars w.r.t. the halo stars, described by $f_{\mathrm{sgr}}$, was then successively lowered; i.e. the fit was carried out in the limit of many and few stars in each  $\tilde{\Lambda}_{\odot}$ slice to make sure that reasonable fits can be obtained for densities like the ones present for the PS1 3$\pi$ RR Lyrae candidates which is ${\sim}$0.5 -- 1 deg$^{-2}$ for most parts of the sky.

\subsection{Fits to Individual $\tilde{\Lambda}_{\odot}$ Bins}

\begin{figure*}
\begin{center}  
\includegraphics[]{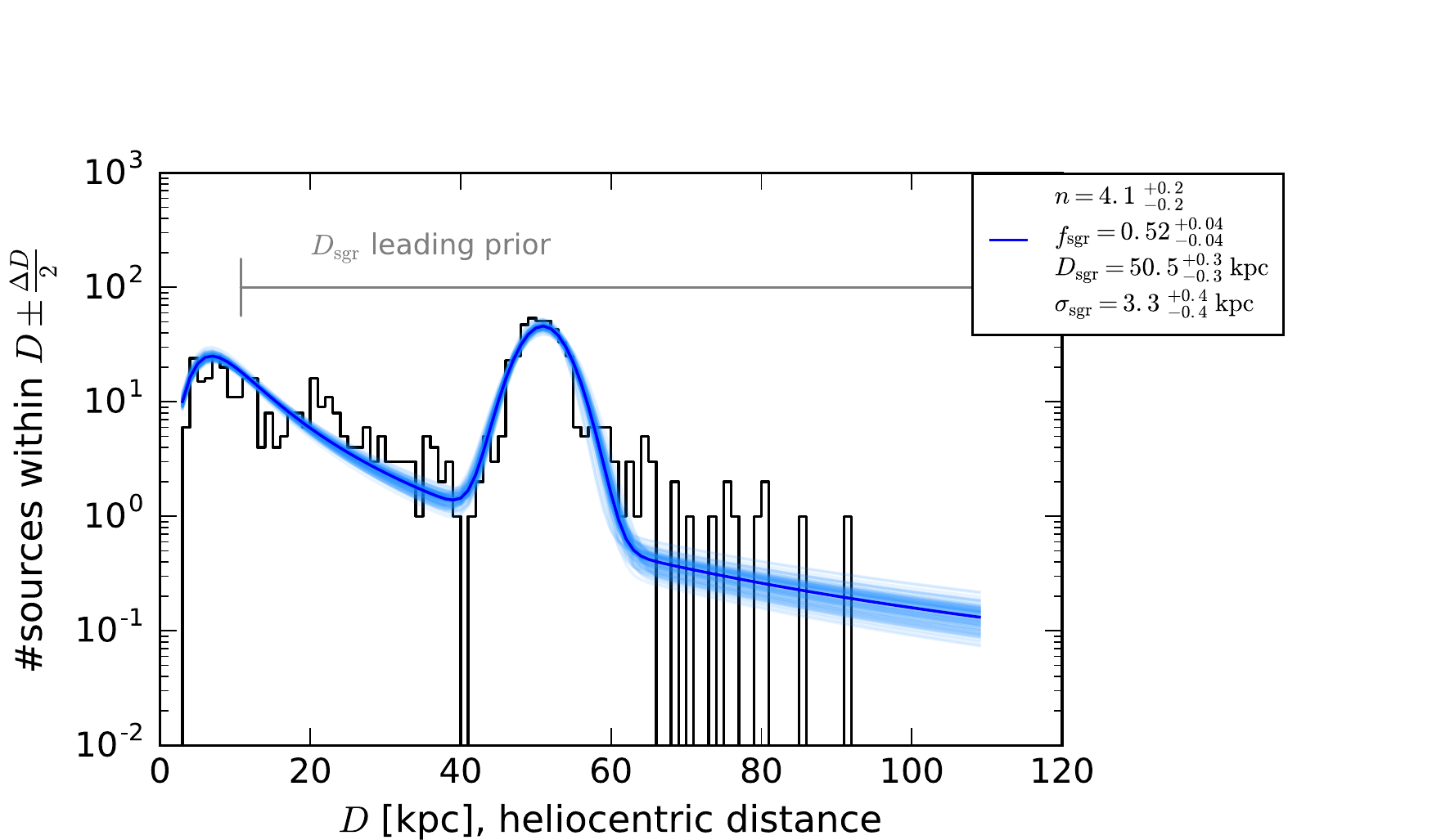}
\caption[Combined halo and stream fit for a $10\arcdeg$ wide slice centered on $\tilde{\Lambda}_{\odot}=55\arcdeg$ where only the leading arm of the Sgr stream is present.]
{{Combined halo and stream fit for a $10\arcdeg$ wide slice centered on $\tilde{\Lambda}_{\odot}=55\arcdeg$.
In this slice, only the leading arm of the Sgr stream is present.\newline
The source distance distribution is shown, overplotted with the model from the best-fit parameters given as solid blue line. The spread of transparent blue lines gives the spread of models obtained by the MCMC. The best-fit parameters are given along with their 1$\sigma$ uncertainties. The plot indicates the prior on $D_{\mathrm{sgr}}$, set by the minimum and maximum distance available from sources in this $\tilde{\Lambda}_{\odot}$ slice.
}
\label{fig:individual_hist_leading}}
\end{center}
\end{figure*}

\begin{figure*}
\begin{center}  
\includegraphics[]{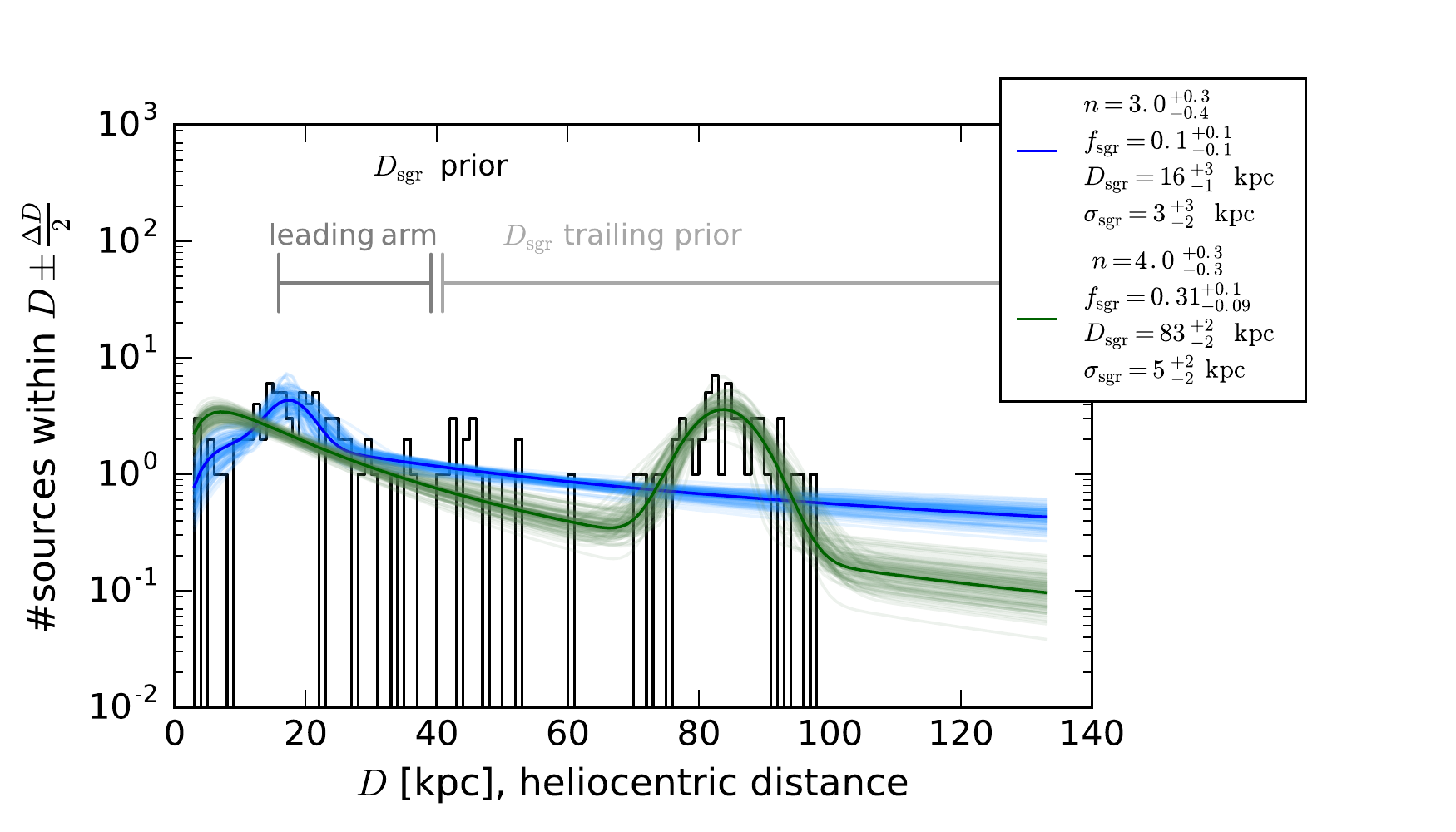}
\caption[Combined halo and stream fit for a $10\arcdeg$ wide slice centered on $\tilde{\Lambda}_{\odot}=155\arcdeg$ where both the leading and trailing arm of the Sgr stream are present.]
{{Combined halo and stream fit for a $10\arcdeg$ wide slice centered on $\tilde{\Lambda}_{\odot}=155\arcdeg$ where both the leading and trailing arm of the Sgr stream are present. For this plot, the fitting was executed twice, with the different priors indicated. The figure is similar to Fig. \ref{fig:individual_hist_leading}, but shows the influence of a carefully chosen prior to separate both debris. Using distinct priors on $D_{\mathrm{sgr}}$, precise estimates on distance and depth of both leading and trailing arm are possible.
}
\label{fig:individual_hist_leading_trailing}}
\end{center}
\end{figure*}

We now illustrate which practical issues are entailed in fitting the model to the data in a $\tilde{\Lambda}_{\odot}$ bin.
Each distance and depth estimate $(D_{\mathrm{sgr}},\sigma_{\mathrm{sgr}})$ is obtained by optimizing Equ. \eqref{equ:evaluate} using a MCMC \citep{Foreman2012}.
Figures \ref{fig:individual_hist_leading} and \ref{fig:individual_hist_leading_trailing} show fits to individual slices in $\tilde{\Lambda}_{\odot}$. 
Fig. \ref{fig:individual_hist_leading} gives the fit for a $10\arcdeg$ wide slice centered on $\tilde{\Lambda}_{\odot}=50\arcdeg$. In these direction, only the leading arm is present. The plot indicates the prior on $D_{\mathrm{sgr}}$, in these cases, only set by the minimum and maximum distance available from sources in the $\tilde{\Lambda}_{\odot}$ slice in case. The distribution of the sources is shown, overplotted with the model from the best-fit parameters given as a solid blue line. The transparent blue lines represent samples drawn from the parameter probability density function,
illustrating the spread of models; the downturn of the models at small distances below $40$~kpc is also a reflection of our sample incompleteness
 (here at the bright end). In the case in Fig. \ref{fig:individual_hist_leading}, showing $\tilde{\Lambda}_{\odot}=55\arcdeg$, a halo profile much steeper than expected from $n=2.62$ given in the \cite{Sesar2013b} model is obvious; local variations in $n$ presumably reflect simply the halo-substructure.
The estimate of $D_{\mathrm{sgr}}$ and $\sigma_{\mathrm{sgr}}$ is clearly seen as being sensible in Fig. \ref{fig:individual_hist_leading}. Here, the variance on the estimated parameters is very small, and the parameters fit well to what one would guess by visual inspection.
Even for $\tilde{\Lambda}_{\odot}$ slices where the fit is poorer (both by visual inspection, and by the variance of the distance estimate)
a sensible distance estimate, not driven by the priors, is found as we show below.

Fig. \ref{fig:individual_hist_leading_trailing} gives the fit for $\tilde{\Lambda}_{\odot}=155\arcdeg$, where both leading and trailing arms are along the line of sight. Using distinct priors on $D_{\mathrm{sgr}}$, separates both debris and gives precise estimates on distance and depth of both leading and trailing arm (see also Fig. \ref{fig:bestfit} around $\tilde{\Lambda}_{\odot}=155\arcdeg$). This illustrates the importance of carefully set priors.

\section{Results}
\label{sec:results}

\begin{figure*}
\begin{center}  
\subfigure[ ]{\includegraphics[]{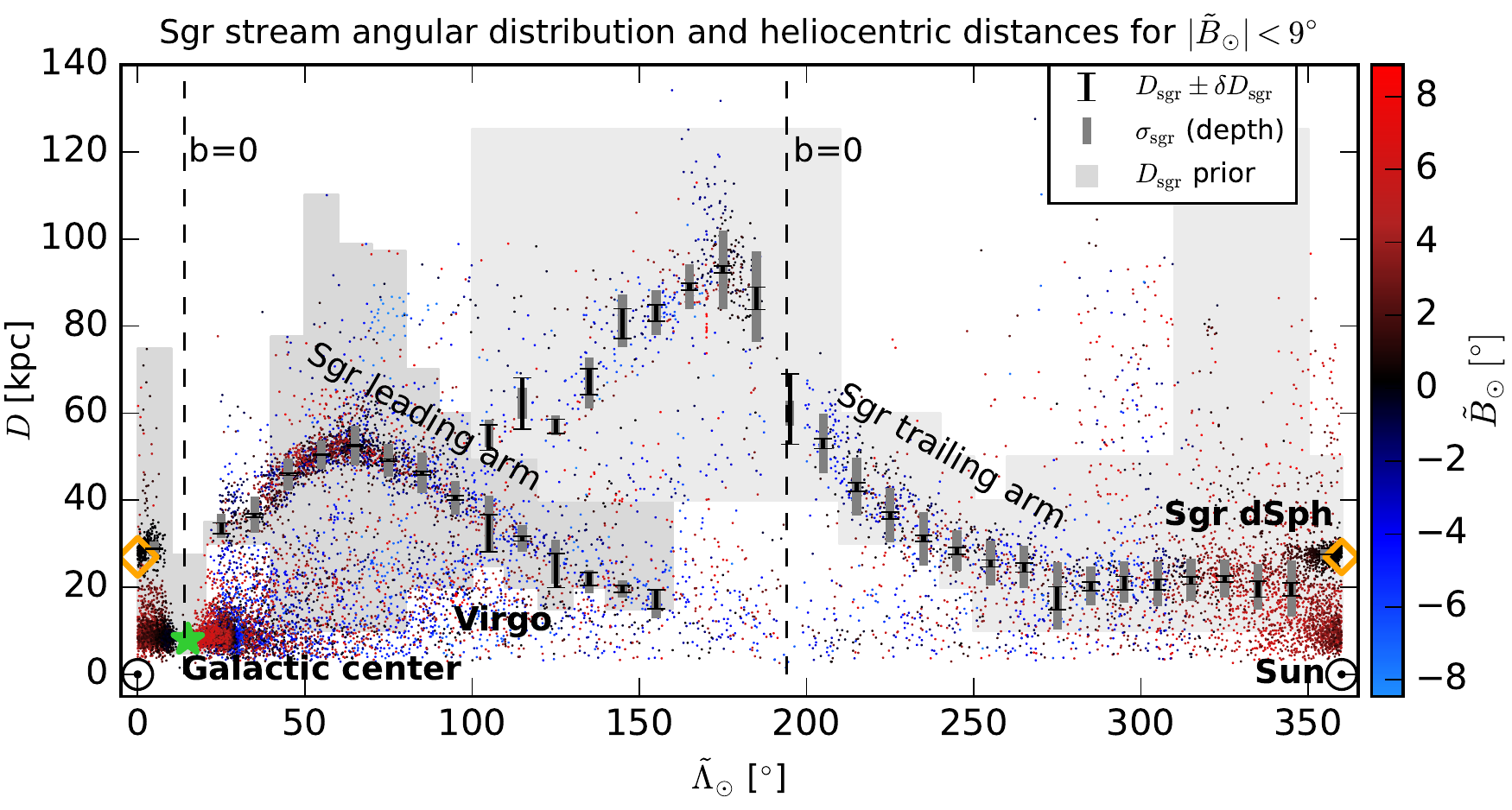}} 
\subfigure[ ]{\includegraphics[trim=0.0inch 0.18inch 0.1inch 0.12inch, clip=true,width=1.1\textwidth]{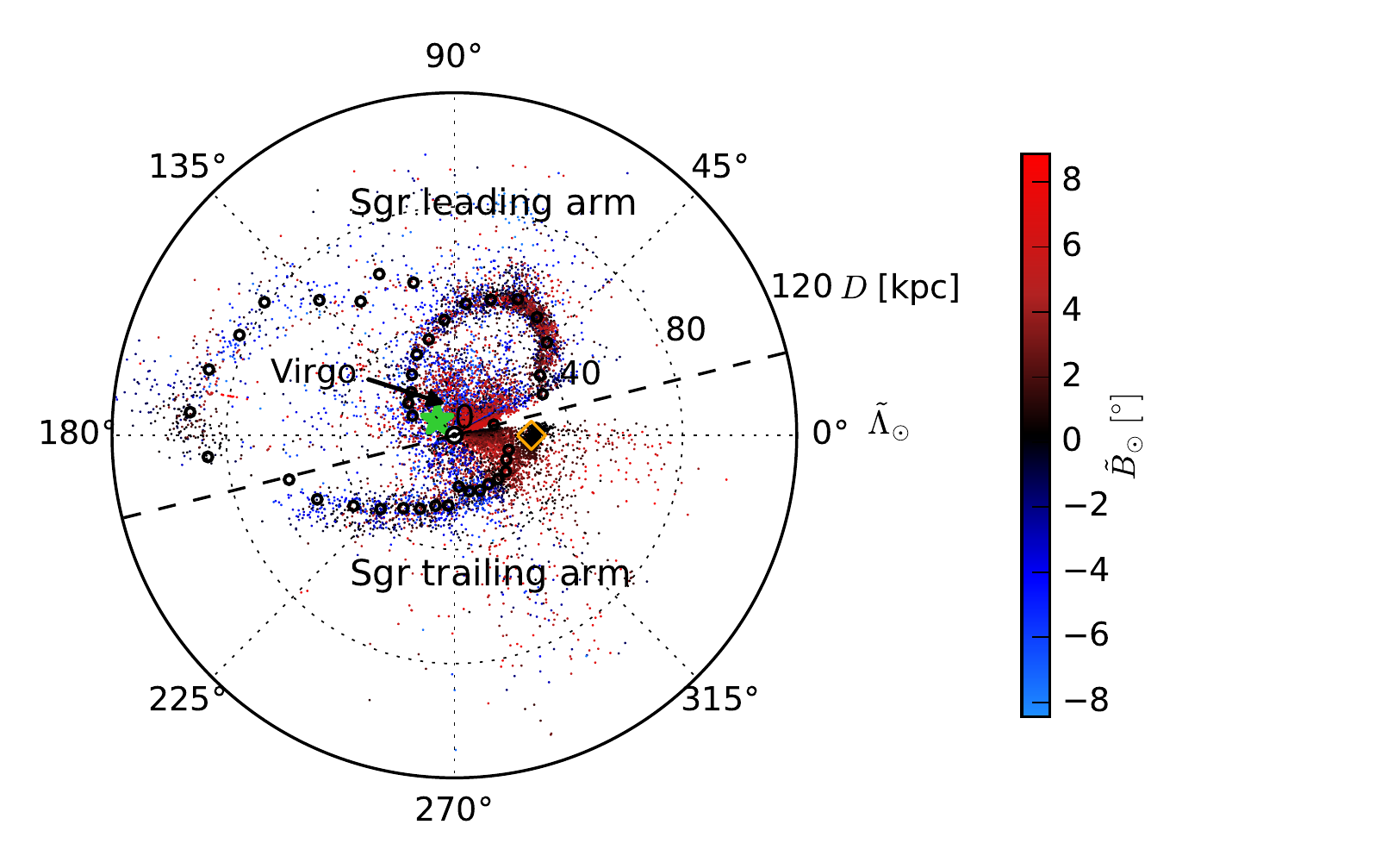}} 
\caption[The extent of the Sgr stream from the RR Lyrae candidates within $\pm 9\arcdeg$ of the Sagittarius plane, shown in Sagittarius coordinates from \cite{Belokurov2014}.]
{{The source distance distribution is shown with the same color coding and symbols as in Fig. \ref{fig:sgr_plot_candidates_sesar_for_paper}, overplotted with the fitted extent of the Sgr stream obtained by the method presented in Sec. \ref{sec:Fitting}.
      (a) The extent of the Sgr stream from the RR Lyrae candidates within $\pm 9\arcdeg$ of the Sagittarius plane, shown in Sagittarius coordinates from \cite{Belokurov2014}. The best fit model, given by $D_{\mathrm{sgr}}$, $\sigma_{\mathrm{sgr}}$ as obtained for $10\arcdeg$ slices in $\tilde{\Lambda}_{\odot}$, is overplotted. The angular distance of the sources to the Sgr plane $\tilde{B}_{\odot}=0\arcdeg$ is indicated by color-coding. The location of the Sun, Galactic anticenter, Sgr dSph and the Virgo overdensity are indicated. The dashed line marks the position of the Galactic plane. The black points indicate the center of the  $\tilde{\Lambda}_{\odot}$ slices used to estimate the distance 
 $D_{\mathrm{sgr}}$. (b) Projection of the stream and distance model fit in cylindrical coordinates centered on the Sun. The same data, symbols, and color coding apply as in (a).
}
\label{fig:bestfit}}
\end{center}  
\end{figure*}

The modelling from Section \ref{sec:TheModelSgr} was then applied to the complete sample of RRab stars within $\vert \tilde{B}_{\odot} \vert <9\arcdeg$. In Fig. \ref{fig:bestfit}, the resulting geometric characterization of the Sgr stream is shown, its fitted distance and line-of-sight (l.o.s) depth (actually $2\times \sigma_{\mathrm{sgr}}$). It is apparent that the distance and l.o.s. depth estimates trace the stream well all the way out to more than 100~kpc. From this detailed picture of the Sgr stream, many features can be seen in great detail, some of them reported previously.
The distances $D_{\mathrm{sgr}}$ are shown as black points centered on the $\tilde{\Lambda}_{\odot}$ slice in case. Its l.o.s. depth $\sigma_{\mathrm{sgr}}$ is indicated by black bars.
The grey shaded areas mark the priors set on $D_{\mathrm{sgr}}$; clearly in most cases the priors play no significant role for the probability density function.
The fitted parameter values are given in Tab. \ref{tab:sgr_leading_arm} and \ref{tab:sgr_trailing_arm} in the Table Appendix.

Qualitatively the Sgr stream aspects shown in Figure \ref{fig:bestfit} can be summarized as follows:
\begin{compactenum}[(i)]
\item The stream shows clearly distinct leading and trailing arms. The shape and extent look similar to what was found earlier \citep{Majewski2003, Belokurov2014}, see also Section \ref{sec:comparebelokurov}.
\item The leading arm's apocenter lies between $\tilde{\Lambda}_{\odot}=60\arcdeg$ and $70\arcdeg$ where $D_{\mathrm{sgr}}$ reaches $48.5-49.6$~kpc, and the trailing arm's apocenter is near $\tilde{\Lambda}_{\odot}{\sim}170\arcdeg$ reaching its largest extent of 92.0~kpc.
This agrees with \cite{Belokurov2014}, who give the leading arm's apocenter being located at $\tilde{\Lambda}_{\odot}=71\arcdeg.3 \pm 3\arcdeg.3$ and the trailing arm's apocenter at $\tilde{\Lambda}_{\odot}=170\arcdeg.5 \pm 1\arcdeg$. The precise position of the apocenters will be derived in Sec. \ref{sec:apocenters}.
\item At both the apocenters of the main leading arm
($\tilde{\Lambda}_{\odot}{\sim}70\arcdeg$) and trailing arm ($\tilde{\Lambda}_{\odot}{\sim}180\arcdeg$) our RRab map reveals substructure that is readily apparent to the eye and has been more discussed in \cite{Sesar2017b} and \cite{Sesar2017c}: two ``clumps'' (at $D{\sim} 60$ and 80~kpc) beyond the 
leading arm's apocenter, and a ``spur" of the trailing arm reaching up to 130~kpc. Such features were previously predicted by dynamical models of the stream \citep[e.g.][]{ Gibbons2014}. These new Sgr stream features are discussed in \cite{Sesar2017b} in detail.
\end{compactenum}

\subsection{The Line-of-Sight Depth of the Sagittarius Stream}
\label{sec:TheDepthOfTheSagittariusStream}
\begin{figure*}
\begin{center}  
\includegraphics[]{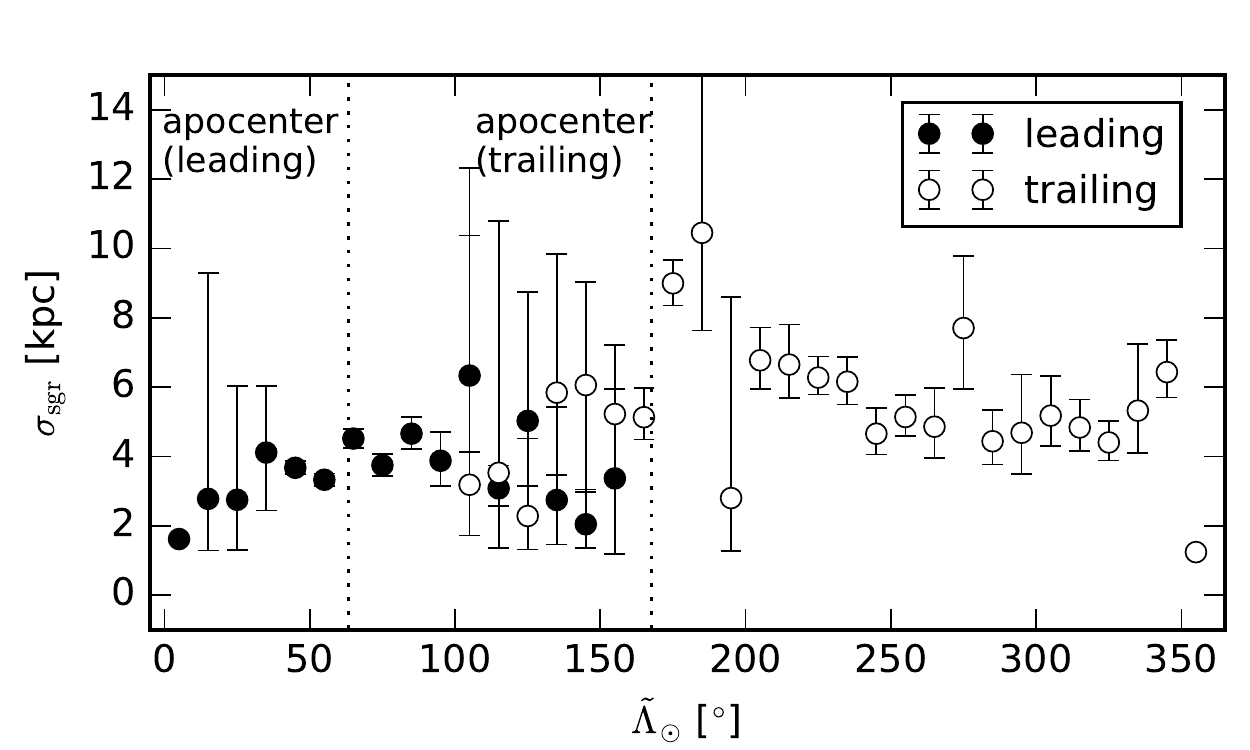}
\caption[The depth $\sigma_{\mathrm{sgr}}$ of the Sagittarius stream from the RRab stars within $\pm 9\arcdeg$ of the Sagittarius plane.]{{
      The depth $\sigma_{\mathrm{sgr}}$ of the Sagittarius stream from the RRab stars within $\pm 9\arcdeg$ of the Sagittarius plane. Error bars indicate the $D_{\mathrm{sgr}}\pm \delta D_{\mathrm{sgr}}$ range. A trend in the depth can be seen, reaching maximum around the apocenters and towards the largest $\tilde{\Lambda}_{\odot}$ of each the leading and trailing arm, respectively.
We find the leading arm's apocenter at $\tilde{\Lambda}_{\odot}^L = 63\arcdeg .2\pm 1\arcdeg .2$, and the trailing arm's apocenter at $\tilde{\Lambda}_{\odot}^T = 167\arcdeg.58\pm 0\arcdeg.44$. The apocenter positions are indicated here as dashed lines.}
\label{fig:widthofstream}}
\end{center}
\end{figure*}

Fig. \ref{fig:widthofstream} shows the estimated $\sigma_{\mathrm{sgr}}$ of the stream (being half the l.o.s. depth) vs. $\tilde{\Lambda}_{\odot}$ for both the leading and trailing arm along with its uncertainty. Fig. \ref{fig:widthofstream} quantifies what was qualitatively apparent from Fig. \ref{fig:bestfit}(a): the stream tends to broaden along its orbit from ${\sim}$1.75~kpc to 6~kpc for the leading arm, reaching even ${\sim}$10~kpc for the trailing arm. As expected, $\sigma_{\mathrm{sgr}}$ and thus the l.o.s. depth is largest close to the apocenters. This is the first systematic determination of the l.o.s. depth, albeit the uncertainties are still quite large for some parts of the stream.
The leading arm's l.o.s. depth rises (and falls) towards (and away) from the apocenter.
In contrast, $\sigma_{\mathrm{sgr}}$ for the trailing arm remains larger between $200^\circ < \tilde{\Lambda}_{\odot}<300^\circ$.
At least in part, this is presumably because our line-of-sight direction forms a shallower angle with the stream direction, compared to the leading arm.
Except towards the apocenters, $\sigma_{\mathrm{sgr}}$ raises also towards the ``end" (the largest $\tilde{\Lambda}_{\odot}$) of the respective trailing or leading arm.

In addition to the l.o.s. depth of the Sgr stream, the actual depth of the stream would be of great interest. As we know the angle between the normal on the stream, and the line of sight, we could deproject the l.o.s. depth $\sigma_{\mathrm{sgr}}$ to get the actual width of the stream.

First, we convert the polar coordinates of the projected $(   \tilde{\Lambda}_{\odot}, \tilde{B}_{\odot}) $, as shown in Fig. \ref{fig:bestfit}, into their Cartesian counterparts  $(x_\mathrm{sgr}, y_\mathrm{sgr})$. We calculate then the deprojected depth $\tilde{\sigma}_{\mathrm{sgr}}$ for each bin $i$ in $\tilde{\Lambda}_{\odot}$ as
\begin{equation}
\tilde{\sigma}_{\mathrm{sgr},i} = \sigma_{\mathrm{sgr},i}\cos \left( \tilde{\Lambda}_{\odot,i} -\alpha_i   \right)
\label{equ:deproject}
\end{equation}   
with
\begin{equation}
\alpha_i = \tan \left(  \frac{y_{\mathrm{sgr},i+1} - y_{\mathrm{sgr},i-1}}{x_{\mathrm{sgr},i+1} - x_{\mathrm{sgr},i-1}} \right).
\label{equ:alpha}
\end{equation}
Equ. \eqref{equ:alpha} approximates the tangent in $(x_{\mathrm{sgr},i},y_{\mathrm{sgr},i})$ with a line through $(x_{\mathrm{sgr},i-1},y_{\mathrm{sgr},i-1})$ and $(x_{\mathrm{sgr},i+1},y_{\mathrm{sgr},i+1})$, thus the first and last $\sigma_{\mathrm{sgr}}$ of the leading and trailing arm are not deprojected.

\begin{figure*}
\begin{center}  
\subfigure[ ]{\includegraphics[trim=0.0inch 0.18inch 0.1inch 0.12inch, clip=true,width=1.1\textwidth]{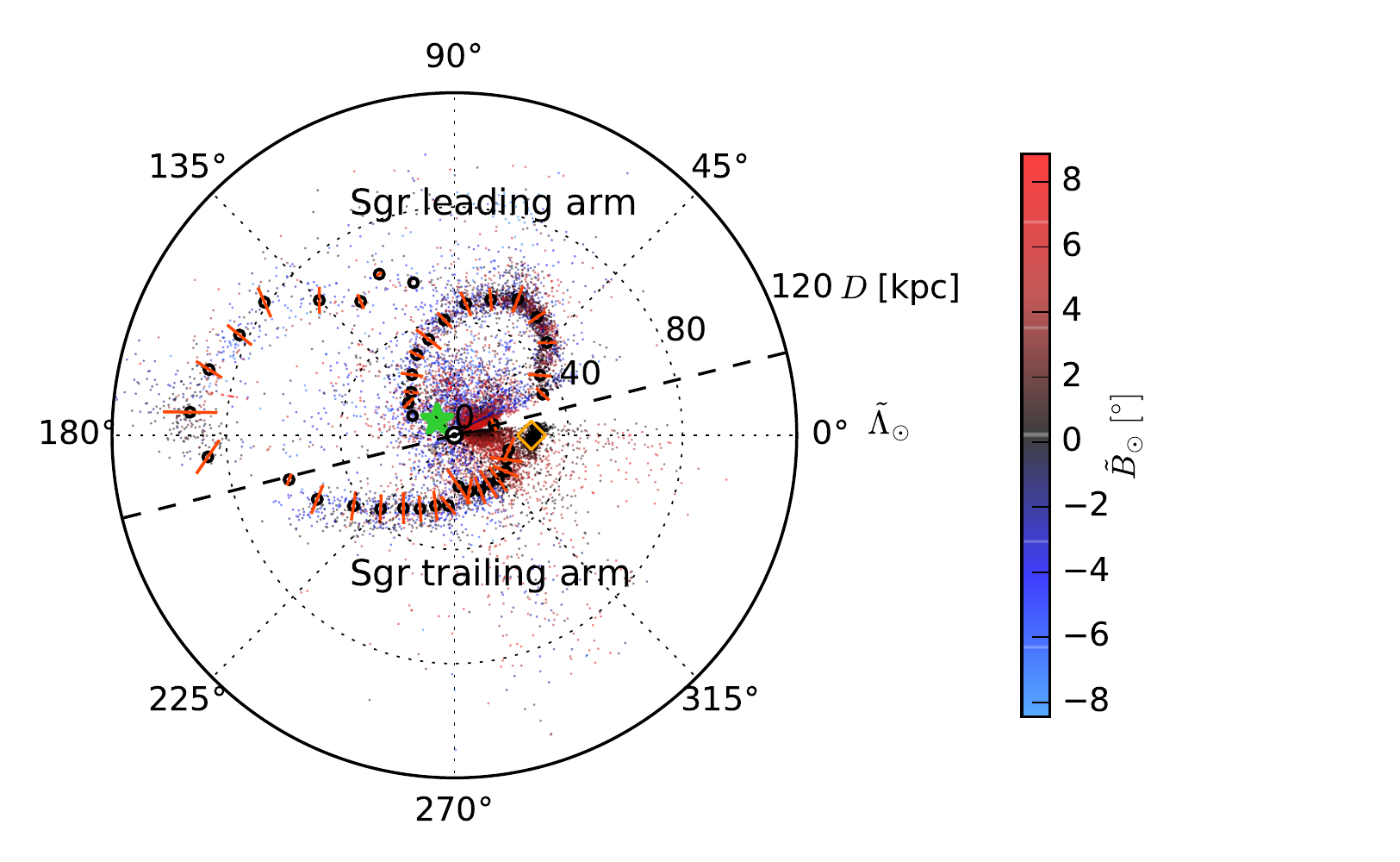}} 
\subfigure[ ]{\includegraphics[]{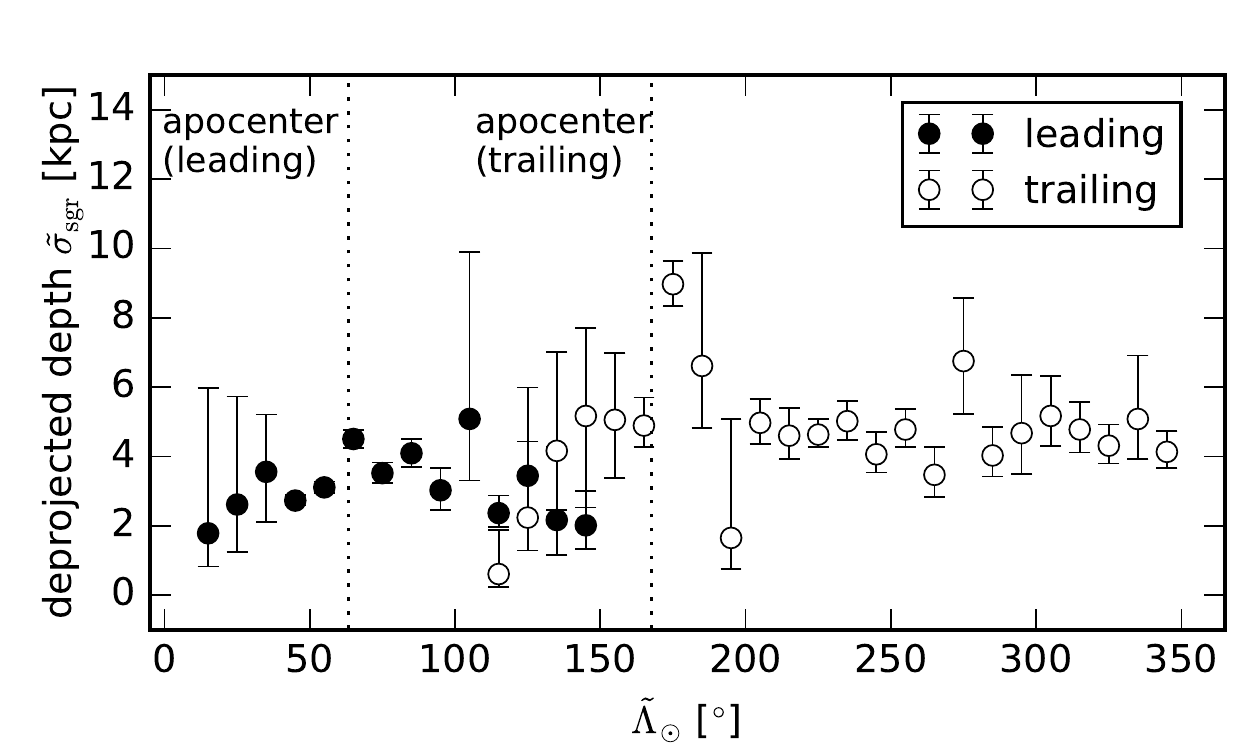}} 
\caption[The deprojected depth of the Sgr stream.]
{{The l.o.s. and deprojected depth of the Sgr stream. \newline
(a) Projection of the stream and its depth in cylindrical Sagittarius coordinates centered on the Sun. The orange bars indicate the deprojected depth, $\tilde{\sigma}_{\mathrm{sgr}}$. The source distance distribution is shown with the same data, symbols and color coding as in Fig. \ref{fig:bestfit}.\newline
(b) The deprojected depth $\tilde{\sigma}_{\mathrm{sgr}}$ of the Sagittarius stream from the RRab stars within $\pm 9\arcdeg$ of the Sagittarius plane. Error bars indicate the $D_{\mathrm{sgr}}\pm \delta D_{\mathrm{sgr}}$ range. The general trend in the depth, seen in Fig. \ref{fig:widthofstream} for the l.o.s. depth $\sigma_{\mathrm{sgr}}$, is still present here, but the profile is flatter than the $\sigma_{\mathrm{sgr}}$ profile from Fig. \ref{fig:widthofstream}, as projection effects contribute to broadening near the apocenters. As in Fig. \ref{fig:widthofstream}, the apocenter positions are indicated as dashed lines. \newline
The deprojected depths along with their uncertainties are given in Tab. \ref{tab:deprojected_leading} and \ref{tab:deprojected_trailing} in the Table Appendix.
}
\label{fig:deprojection}}
\end{center}
\end{figure*}

The deprojected depths along with their uncertainties are given in Tab. \ref{tab:deprojected_leading} and \ref{tab:deprojected_trailing} in the Table Appendix.
Fig. \ref{fig:deprojection} shows how the l.o.s. and deprojected depth of the Sgr stream strongly varies during the orbital period. The $\tilde{\sigma}_{\mathrm{sgr}}$ profile is flatter than the $\sigma_{\mathrm{sgr}}$ profile, and as expected, the trailing arm's deprojected depth $\tilde{\sigma}_{\mathrm{sgr}}$ is not noticeable boosted in contrast to  $\sigma_{\mathrm{sgr}}$ which is. But a variation during the orbital period is still present.
Comparing both depths emphasizes that the larger depth at the apocenters is a combination of the projection effect, as well as of the true broadening when the velocities become small near the apocenters.

\subsection{The Amplitude of the Sagittarius Stream}
\label{sec:TheAmplitudeOfTheSagittariusStream}

We can also quantify the amplitude $A$ of the stream fit from Sec. \ref{sec:TheModelSgr}., defined as the number of RRab stars in the stream per degree as a function of $\tilde{\Lambda}_{\odot}$, i.e.:

\begin{equation}
A = \left(\mathrm{\#sources\;within\;}{\tilde{\Lambda}}_{\odot} \pm \frac{\Delta \mathrm{\tilde{\Lambda}}_{\odot}}{2}\right) \times f_\mathrm{sgr} / (\Delta \mathrm{\tilde{\Lambda}_{\odot}} \times \sigma_{\mathrm{sgr}}).
\label{equ:amplitude}
\end{equation}

The amplitudes for both the leading and the trailing arm are given in Tab. \ref{tab:amplitude_weighted_B_leading} and \ref{tab:amplitude_weighted_B_trailing} in the Table Appendix.

\begin{figure*}
\begin{center}  
\includegraphics[]{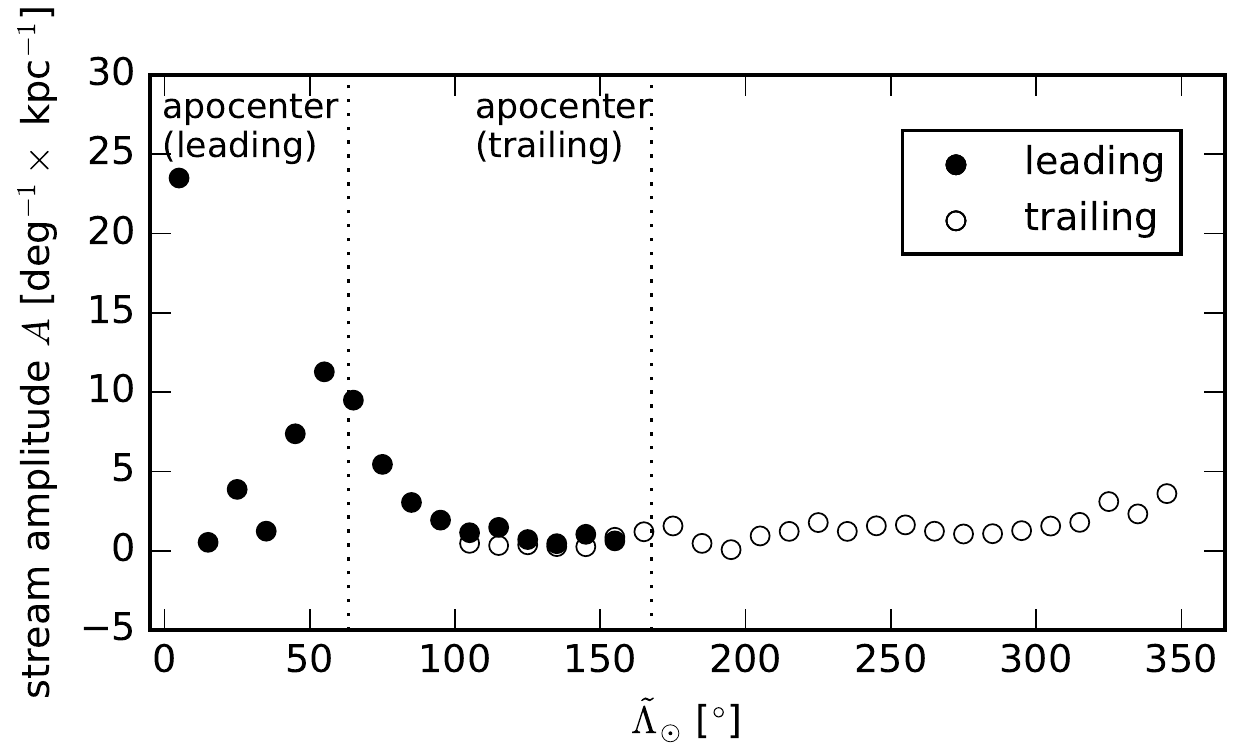}
\caption[The amplitude of the stream.]{{
      The amplitude of the stream, $A = (\mathrm{\#sources\;within\;}{\tilde{\Lambda}}_{\odot} \pm \frac{\Delta \mathrm{\tilde{\Lambda}}_{\odot}}{2}) \times f_\mathrm{sgr} / (\Delta \mathrm{\tilde{\Lambda}_{\odot}} \times \sigma_{\mathrm{sgr}})$, for the $\tilde{\Lambda}_{\odot}$ bins. $A$ shows an increase towards the apocenters, and is about six times larger near the leading arm's apocenter than near the trailing arm's. As in previous figures, the apocenter positions are indicated as dashed lines.}
\label{fig:sgr_fit_amplitude}}
\end{center}
\end{figure*}

Fig. \ref{fig:sgr_fit_amplitude} shows the amplitudes plotted vs. the $\tilde{\Lambda}_{\odot}$ bins. The value of $A$ increases near the apocenter of the leading arm to about twice as much as away from its apocenter. Also near the apocenter of the trailing arm, $A$ rises w.r.t. the value it has away from the apocenter, but not as striking as found for the leading arm. As the angular velocity decreases near the apocenter, we had expected finding an increased source density, and thus larger $A$, near the apocenters compared to sections of the stream away from apocenters. In addition to this general statement, we find that the source density is about six times larger at the leading arm's than at the trailing arm's apocenter (compare also Fig. \ref{fig:sgr_fit_amplitude} to Fig. \ref{fig:bestfit}). 
We checked if this can be partially explained by a selection effect, as the leading arm's apocenter has a smaller heliocentric distance than the trailing arm's. Using the selection function from \cite{Sesar2017b}, we find that this will by far not account for the difference between the leanding and trailing arm's apocenter source densities, so incompleteness is not an issue here. Additionally, simulations like \cite{Dierickx2017} also show a similar behavior, see e.g. \cite{Dierickx2017} Fig. 9 which shows a higher source density at the leading arm's apocenter.

\subsection{The Apocenters and Orbital Precession of the Sagittarius Stream}
\label{sec:apocenters}

Sources orbiting in a potential show a precession of their orbits, which means that they do not follow an identical orbit each time, but actually trace out a shape made up of rotated orbits. This is because the major axis of each orbit is rotating gradually within the orbital plane.

Orbits in the outer regions of galaxies with a spherically symmetric gravitational potential are expected to have a precession between $0\arcdeg$ and $120\arcdeg$ \citep{Belokurov2014}. Assuming a spherically
symmetric potential, the precession depends primarily on the shape of the potential and thus the radial mass distribution \citep{Belokurov2014}. Additionally it is also a function of the orbital energy and angular momentum distribution \citep{Binney2008}.

The angular mean distance estimates $D_{\mathrm{sgr}}$ of the Sgr stream that were obtained during this work enable us to make statements about the precession of the orbit. For doing so, the angle between the leading and the trailing apocenters is measured.

We calculate this angle by fitting a model to the distance data in both the leading and trailing arm, namely fitting a Gaussian and a (shifted and scaled) log-normal. A comparable fit was carried out by \cite{Belokurov2014}. The models used here are unphysical, but can be applied here as they describe the angular distance distribution $D_{\mathrm{sgr}} (\tilde{\Lambda}_{\odot})$ adequately in order to find the apocenters along with their uncertainties. As the angular distance distribution $D_{\mathrm{sgr}} (\tilde{\Lambda}_{\odot})$ of the leading arm appears to be symmetrical w.r.t. to the assumed apocenters, as well as appears to be Gaussian-like, a Gaussian model is fitted to the $D_{\mathrm{sgr}} \tilde{\Lambda}_{\odot})$ of the leading arm. In contrast, the trailing arm shows a clear asymmetry. fort this reason, we fit the trailing arm's distance distribution using a (shifted an scaled) log-normal, fitted for the range $105\arcdeg \leq \tilde{\Lambda}_{\odot} \leq 265\arcdeg$. With comparable results, a parabola can be fitted to the data. 

\begin{figure*}
\begin{center}  
\includegraphics[]{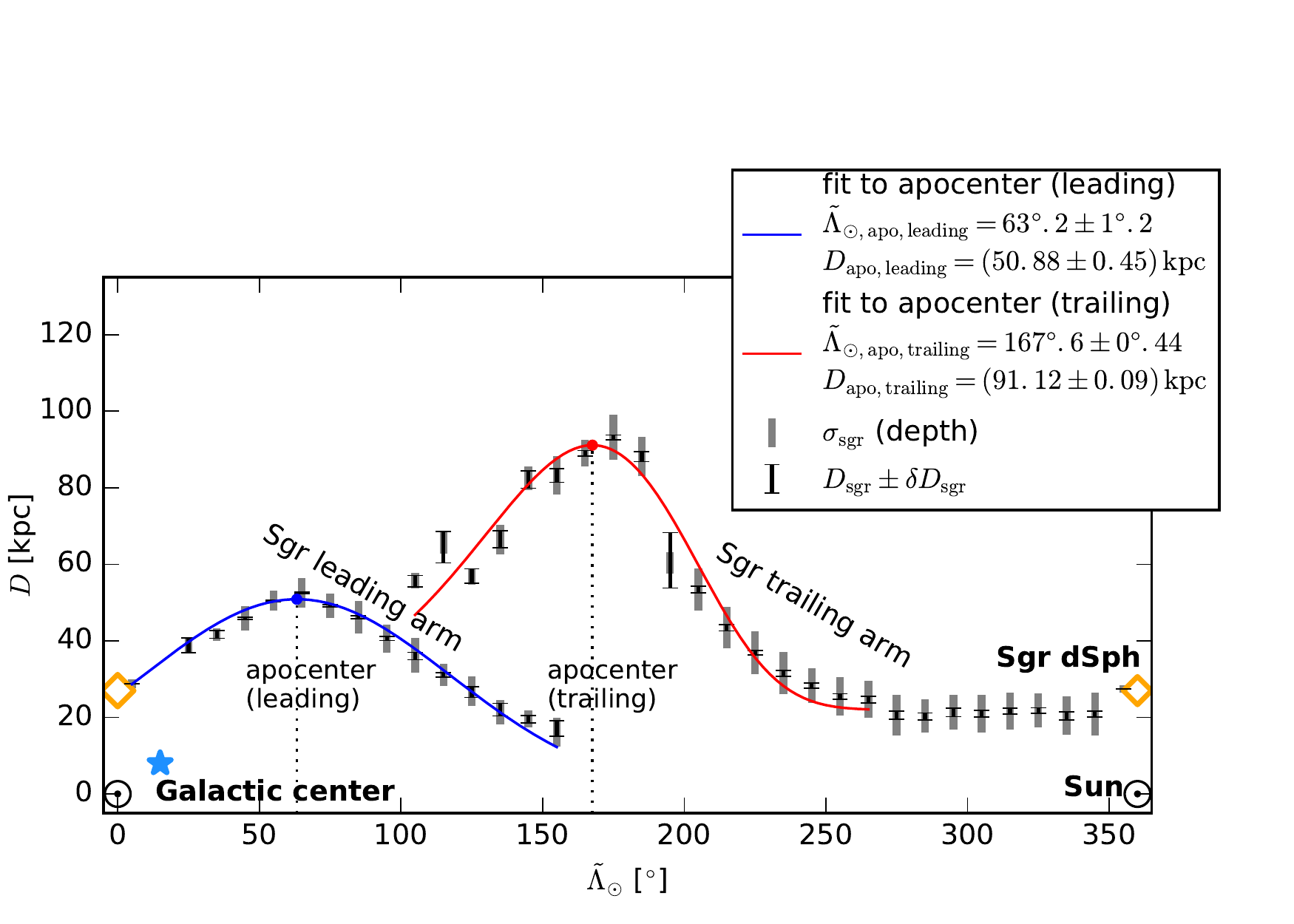}
\caption[The apocenters of Sgr stream.]{{The apocenters of Sgr stream by fitting $D_{\mathrm{sgr}} (\tilde{\Lambda}_{\odot})$ with a Gaussian for the leading arm and a lognormal to the trailing arm.
We derive the position of the leading apocenter as $\tilde{\Lambda}_{\odot}^L = 63\arcdeg .2\pm 1\arcdeg .2$, reaching $D_{\mathrm{sgr}}^L = 50.88 \pm 0.45~\mathrm{kpc}$, and the 
trailing apocenter as $\tilde{\Lambda}_{\odot}^T = 167\arcdeg. 6\pm 0\arcdeg.44$, reaching $D_{\mathrm{sgr}}^T = 91.12 \pm 0.09~\mathrm{kpc}$. From this, we calculate the differential heliocentric orbit precession $\omega_{\odot}=\tilde{\Lambda}_{\odot}^T-\tilde{\Lambda}_{\odot}^L = 104\arcdeg .4 \pm 1\arcdeg .3$. The corresponding difference in heliocentric apocenter distances is $40.24\pm 0.45$~kpc.\newline
Blue and red lines show the best-fit models for both the leading and trailing arm. The position of the apocenters is denoted by a circle 
symbol in the corresponding color. Dashed lines mark the corresponding $\tilde{\Lambda}_{\odot}$ of each apocenter.}
\label{fig:apocenters}}
\end{center}
\end{figure*}

The best-fit Gaussian model for the leading and trailing apocenters, respectively, is shown in Fig. \ref{fig:apocenters}. 

In this figure, blue and red lines show the best-fit Gaussian model for both the leading and trailing arm. The position of the apocenters is denoted by a circle 
symbol each. Dashed lines mark the apocenter's corresponding $\tilde{\Lambda}_{\odot}$.

We find the leading apocenter at $\tilde{\Lambda}_{\odot}^L = 63\arcdeg .2\pm 1\arcdeg .2$, reaching $D_{\mathrm{sgr}}^L = 50.88 \pm 0.45~\mathrm{kpc}$, and the 
trailing apocenter at $\tilde{\Lambda}_{\odot}^T = 167\arcdeg. 6\pm 0\arcdeg.44$, reaching $D_{\mathrm{sgr}}^T = 91.12 \pm 0.09~\mathrm{kpc}$. 

For a more detailed discussion of the apocenter substructure, reaching up to 120 kpc from the Sun, we refer to \cite{Sesar2017b}, Section 3.

The differential orbit precession $\omega_{\odot}=\tilde{\Lambda}_{\odot}^T-\tilde{\Lambda}_{\odot}^L$ is $104\arcdeg .4 \pm 1\arcdeg .3$, corresponding to a difference in heliocentric apocenter distances of $40.24\pm 0.45$~kpc.

The actual Galactocentric orbital precession is slightly lower than the difference between the heliocentric apocenters. 
The Galactocentric distances and angles of the leading and trailing apocenters are calculated by taking into account the Galactocentric distance of the Sun being 8~kpc. 
Consequently, the opening angle between the positions of the two apocenters, as viewed from Galactic center, is then $\omega_{\mathrm{GC}} =  96\arcdeg.8 \pm 1\arcdeg.3$. 
The Galactocentric distance of the leading apocenter is then $47.8 \pm 0.5$~kpc, and of the trailing apocenter $98.95 \pm 1.3$~kpc, resulting into a difference in mean Galactocentric apocenter distances of $47.45\pm1.4$~kpc.

\subsection{The Orbital Plane Precession of the Sagittarius Stream}
\label{sec:TheOrbitalPlanePrecessionoftheSagittariusStream}
Aside from the apocenter precession of the stream (see Sec. \ref{sec:apocenters}), the orbital plane itself might show a precession. To test this we obtain the weighted latitude of the stream RRab, $\langle{\tilde{B}_{\odot}}\rangle$, as a function of $\tilde{\Lambda}_{\odot}$. The weight of each star
is the probability that the star is associated with the Sgr stream.

For each bin $i$ in $\tilde{\Lambda}_{\odot}$, the fit as described in Sec. \ref{sec:Fitting} was carried out, resulting in a parameter set $\boldsymbol{\theta}_i= (f_{\mathrm{sgr},i},D_{\mathrm{sgr},i},\sigma_{\mathrm{sgr},i},n_i)$ describing the stream and halo properties in the $\tilde{\Lambda}_{\odot}$ bin in case.

We now again make use of the model for the observed heliocentric distances, Equ. \eqref{eq:stream_halo_model} with the halo described by Equ. \eqref{eqn:halomodel} and the stream described by Equ. \eqref{eqn:streammodel}.
We calculate $p_{\mathrm{sgr}}(l_j,b_j,D_j|\boldsymbol{\theta}_i)$ as the fraction of the likelihood a star $j$ is associated with the Sgr stream divided by the sum of the likelihood that it is associated with Sgr stream and the likelihood that the star is associated with the halo:

\begin{equation}
p_{\mathrm{sgr},j}(l_j,b_j,D_j|\boldsymbol{\theta}_i) = \frac{p_{\mathrm{stream}}(D|\boldsymbol{\theta}_i) }{ (p_{\mathrm{halo}}(D|\boldsymbol{\theta}_i) + p_{\mathrm{stream}}(D|\boldsymbol{\theta}_i))}
\label{equ:p_sgrmember}
\end{equation}

The weighted latitude $\langle{\tilde{B}_{\odot}}\rangle$ in a bin $i$ is then calculated as 

\begin{equation}
\langle{\tilde{B}_{\odot}}\rangle_i = \frac{  \sum_j ( \tilde{B}_{\odot,i} \times p_{\mathrm{sgr},j} )     }{  \sum_j( p_{\mathrm{sgr},j)}}
\label{equ:weighted_B_o}
\end{equation}

We then use the difference in $\langle{\tilde{B}_{\odot}}\rangle$ for the leading and trailing arm to quantify the orbital plane precession.

\begin{figure*}
\begin{center}  
\includegraphics[]{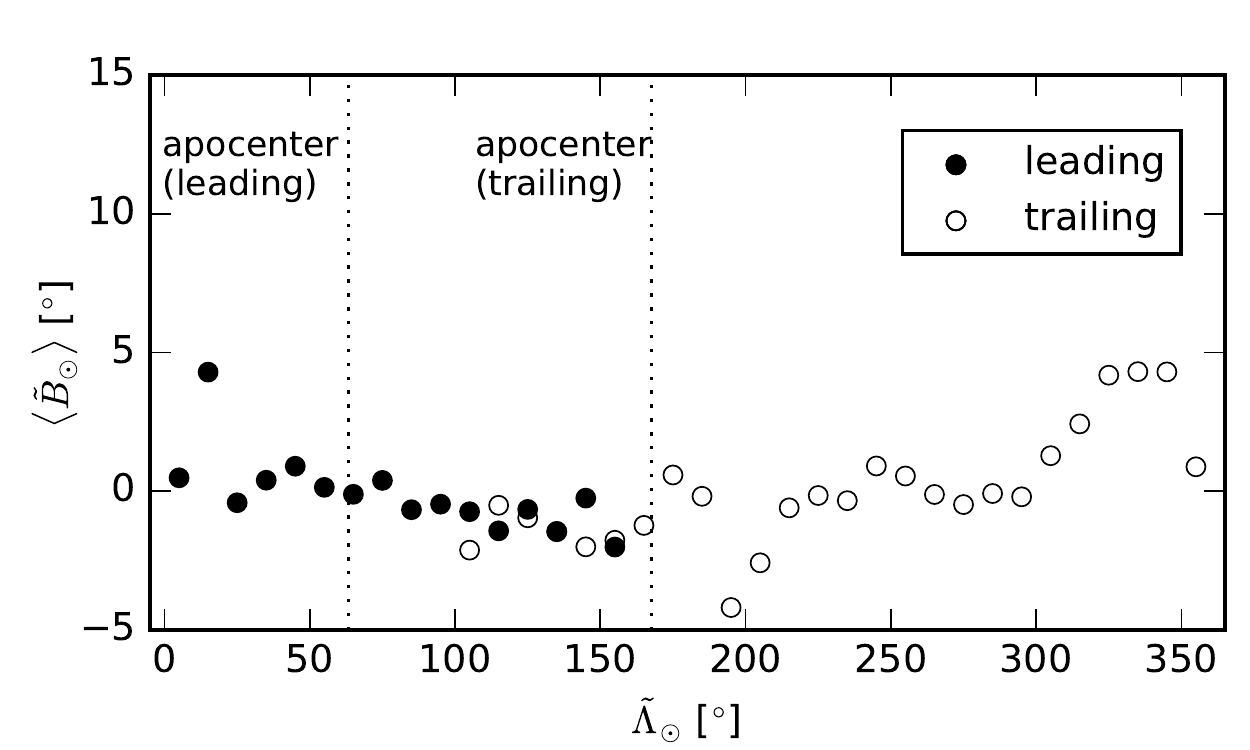}
\caption[The weighted latitude of the stream RRab, $\langle{\tilde{B}_{\odot}}\rangle$, for the $\tilde{\Lambda}_{\odot}$ bins.]{{The weighted latitude of the stream RRab, $\langle{\tilde{B}_{\odot}}\rangle$, for the $\tilde{\Lambda}_{\odot}$ bins. The weight of each star is the probability that the star is associated with the Sgr stream, and  $\langle{\tilde{B}_{\odot}}\rangle$ is then calculated by Equ. \eqref{equ:weighted_B_o}.
Except for $\tilde{\Lambda}_{\odot}= 15 \arcdeg$, the leading arm stays in or is close to the plane defined by ${\tilde{B}_{\odot}}=0\arcdeg$. In contrast, the trailing arm is found within $-5\arcdeg$ to $5\arcdeg$ around the plane. This results into a separation of ${\sim}10\arcdeg$, as also derived by \cite{Law2005a}.}
\label{fig:sgr_weighted_B}}
\end{center}
\end{figure*}

The resulting $\langle\tilde{B}_{\odot}\rangle$ for both the leading and the trailing arm are given in Tab. \ref{tab:amplitude_weighted_B_leading} and \ref{tab:amplitude_weighted_B_trailing} in the Table Appendix. Fig. \ref{fig:sgr_weighted_B} shows $\langle\tilde{B}_{\odot}\rangle$ plotted vs. the $\tilde{\Lambda}_{\odot}$ bins.

This gives evidence for the leading arm staying in or close to the plane defined by ${\tilde{B}_{\odot}}=0\arcdeg$, whereas the trailing arm is found within within $-5\arcdeg$ to $5\arcdeg$ around the plane. From this, we find a separation of ${\sim}10\arcdeg$, as also derived by \cite{Law2005a}.

\section{Discussion}
\label{sec:discussion}

We can now put our results in the context of existing work, and discuss the prospect of using them for dynamical stream modeling.

\subsection{Comparison to the Model by Belokurov et al. (2014)}
\label{sec:comparebelokurov}

\begin{figure*}
\begin{center}  
\includegraphics[]{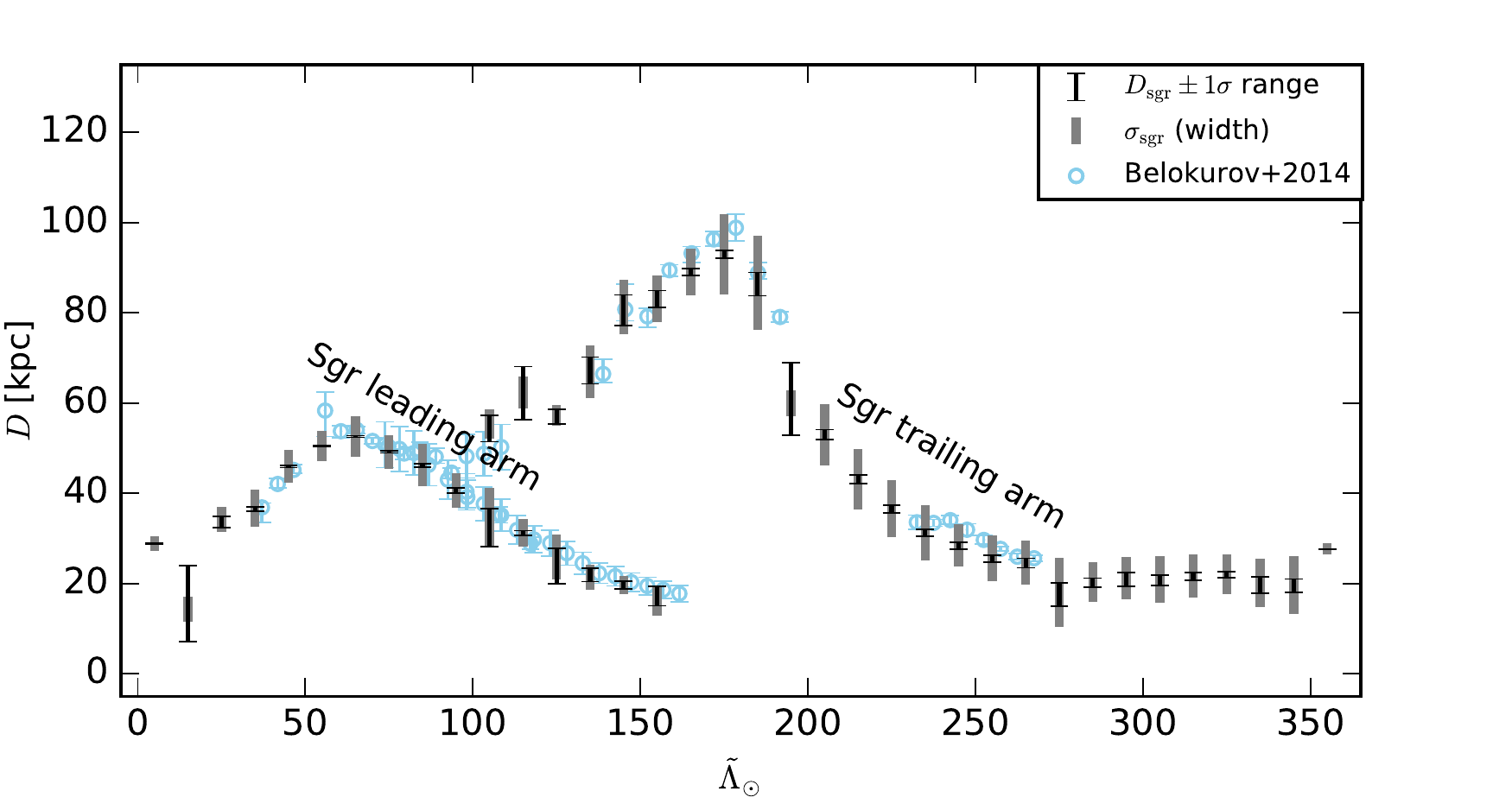}
\caption[Comparison of the heliocentric distance estimates of the Sgr stream between this work and \cite{Belokurov2014}]{{Comparison of the heliocentric distance estimates of the Sgr stream between this work and \cite{Belokurov2014}. The $D_{\mathrm{sgr}}$, shown as black points together with their $D_{\mathrm{sgr}}\pm \delta D_{\mathrm{sgr}}$ range and estimated stream depth $\sigma_{\mathrm{sgr}}$ (grey bars), are compared to the estimates from \cite{Belokurov2014} (blue points) who traced parts of the Sgr stream, together with their uncertainties. The distances from \cite{Belokurov2014} show a slight trend towards larger values. Over all, the distance estimates are in good agreement. \newline
Uncertainties from our results are given as $D_{\mathrm{sgr}}\pm \delta D_{\mathrm{sgr}}$ ranges; uncertainties from \cite{Belokurov2014} are given as their $1\sigma$ ranges if available, and assumed to be $10\%$ if not stated otherwise.} 
\label{fig:sgr_distances_coords_compare_others_distances}}
\end{center}
\end{figure*}

The best previous estimates of the heliocentric distances for a large part of the Sgr stream come from \cite{Belokurov2014}, who used blue horizontal branch (BHB) stars, sub-giant branch (SGB) stars and red giant branch (RGB) stars from the Sloan Digital Sky Survey Data Release 8 (SDSS DR8). 
In Fig.~\ref{fig:sgr_distances_coords_compare_others_distances} we compare our heliocentric distances, $D_{\mathrm{sgr}}\pm \delta D_{\mathrm{sgr}}$ to those from \cite{Belokurov2014} (Figure 6 therein). We show the $1\sigma$ uncertainties from \cite{Belokurov2014} where available, and assume the uncertainties to be $10\%$ if not stated otherwise.

Overall, the two estimates are in good agreement, attesting to the quality of the
\cite{Belokurov2014} analysis. The distances from \cite{Belokurov2014} may be systematically slightly larger; the fact 
that the RRab distances we use are directly tied to HST and Gaia DR1 parallaxes
\citep{Sesar2017b} should lend confidence to the distance scale of this work.
Our new estimates for the mean distance are three times more precise, and presumably also accurate.
The typical mean distance uncertainty in \cite{Belokurov2014} is $1-2$~kpc and up to 0.1 $D_{\mathrm{sgr}}$ for most parts of the stream, whereas our work shows comparable or smaller $\delta D_{\mathrm{sgr}}$ (see Tables \ref{tab:sgr_leading_arm}, \ref{tab:sgr_trailing_arm}).

As mentioned before, our new Sgr stream map also has considerably more extensive angular coverage. 

The high individual distance precision to the RRab of $3\%$ allows us to map the l.o.s. depth of the stream, which \cite{Belokurov2014} could not do, or at least did not. For these reasons, our work improves the knowledge on the geometry of Sgr stream significantly.

However, care must be taken in parts of the Sgr stream where the number of sources is comparably low. The trailing arm's distance estimate for the bin centered on $\tilde{\Lambda}_{\odot}=125 \arcdeg$ results from only 28 sources within the prior indicated by Fig. \ref{fig:bestfit}(a), i.e. $D>40$~kpc. In this bin, the estimated $D_{\mathrm{sgr}}$ is smaller than the $D_{\mathrm{sgr}}$ estimated for nearby bins, and the same applies for the estimated width of the stream which appears being too tight.

In both analyses, the apocenters of the leading and trailing arms are derived.

We find the leading apocenter at $\tilde{\Lambda}_{\odot}^L = 63\arcdeg .2\pm 1\arcdeg .2$, reaching $D_{\mathrm{sgr}}^L = 50.88 \pm 0.45~\mathrm{kpc}$, and the 
trailing apocenter at $\tilde{\Lambda}_{\odot}^T = 167\arcdeg .6\pm 0\arcdeg.44$, reaching $D_{\mathrm{sgr}}^T = 91.12 \pm 0.09~\mathrm{kpc}$. 
The differential orbit precession $\omega_{\odot}=\tilde{\Lambda}_{\odot}^T-\tilde{\Lambda}_{\odot}^L$ is $104\arcdeg .4 \pm 1\arcdeg .3$, with a difference in heliocentric apocenter distances of $40.24\pm 0.45$~kpc.
Taking into account the Galactocentric distance of the Sun being 8~kpc, the corresponding Galactocentric angle from our analysis is $\omega_{\mathrm{GC}} =  96\arcdeg.8 \pm 1\arcdeg.3$. 
The Galactocentric distance of the leading apocenter is then $47.8 \pm 0.5$~kpc, and of the trailing apocenter $98.95 \pm 1.3$~kpc, resulting into a difference in mean Galactocentric apocenter distances of $47.45\pm1.4$~kpc.

If we would assume the trailing arm's apocenter is close to the maximum extent of the derived $D_{\mathrm{sgr}}$, as done by fitting a Gaussian to the five closest points near the maximum extent, we find trailing apocenter at $\tilde{\Lambda}_{\odot}^T = 173\arcdeg 4\pm 2\arcdeg.0$, reaching $D_{\mathrm{sgr}}^T = 92.7 \pm 1.3~\mathrm{kpc}$.
The differential orbit precession $\omega_{\odot}=\tilde{\Lambda}_{\odot}^T-\tilde{\Lambda}_{\odot}^L$ is then $108.9 \pm 2.4\arcdeg$, with a difference in heliocentric apocenter distances of $41.82\pm 0.45$~kpc. The Galactocentric angle is then slightly larger than for the log-normal fit, $\omega_{\mathrm{GC}} =  101\arcdeg.0 \pm 2\arcdeg.4$, the Galactocentric distance of the leading apocenter $49.2 \pm 0.5$~kpc, to the trailing apocenter $100.7 \pm 1.3$~kpc, resulting into a difference in mean Galactocentric apocenter distances of $51.5\pm1.4$~kpc.

\cite{Belokurov2014} give the position of the leading apocenter as $\tilde{\Lambda}_{\odot}^L = 71\arcdeg .3\pm 3\arcdeg .5$ with a Galactocentric distance $R^L=47.8 \pm 0.5$~kpc, and the position of the trailing apocenter as $\tilde{\Lambda}_{\odot}^L = 170\arcdeg .5\pm 1\arcdeg$ with a Galactocentric distance $R^L=102.5 \pm 2.5$~kpc. They state the derived Galactocentric orbital precession as $\omega = 93\arcdeg.2 \pm 3\arcdeg .5$. \\

To summarize the comparison: 

\begin{itemize}
\item Our analysis is done from one single survey and type of stars, whereas the work by \cite{Belokurov2014} relies on BHB, SGB and RGB stars. The extent and depth of PS1 3$\pi$ enables us to provide a more extensive angular coverage of sources. This resulted into the first complete (i.e., spanning $0\arcdeg<\tilde{\Lambda}_{\odot}<360\arcdeg$) trace of Sgr stream's heliocentric distance from a single type of stars originating from a single survey.
\item The heliocentric mean distances of the stream as from \cite{Belokurov2014} may be systematically slightly larger; the fact 
that the RRab distances we use are directly tied to HST and Gaia DR1 parallaxes \citep{Sesar2017b} should lend confidence to the distance scale of this work.
\item Along with the extent of the Sgr stream, we can give its l.o.s. depth $\sigma_{\mathrm{sgr}}$, and deproject $\sigma_{\mathrm{sgr}}$ in order to get its true width.
\item Our analysis shows a Galactocentric orbital precession being about $4 \arcdeg$ larger than as measured by \cite{Belokurov2014}, or $8 \arcdeg$ larger if assuming the trailing arm's apocenter is close to the maximum extent of the derived $D_{\mathrm{sgr}}$.
This is within the error range given by \cite{Belokurov2014}. Generally speaking, the higher the Galactocentric orbital precession, 
the smoother the dark matter densitiy is as a function of the Galactocentric radius. Logarithmic haloes should show an orbital 
precession of about $120 \arcdeg$ \citep{Belokurov2014}, whereas a smaller orbital precession angle indicats a profile with a sharper 
drop in the radial dark matter density \citep{Belokurov2014}.
Finding this result, together with the result of \citep{Belokurov2014} as well as the simulation by \cite{Dierickx2017} 
is a strong indicator that a steeper profile than the logarithmic one should be considered for the dark matter halo of the Milky Way.
\end{itemize}

\subsection{Bifurcation of the Leading Arm}
\label{sec:BifurcatioOfTheLeadingArm}
Part of the Sgr stream's leading arm in the Galactic Norther hemisphere is ``bifurcated'',
or branched, in its projection on the sky \citep{Belokurov2006}. Starting at $\mathrm{RA} \sim 190\arcdeg$, the lower and upper declination branches of the stream, labeled A and B respectively \citep{Belokurov2006}, can be traced at least until  $\mathrm{RA} \sim 140\arcdeg$. As stated by \cite{Fellhauer2006}, the bifurcation likely arises from different stripping epochs, the young leading arm providing branch A and the old trailing arm branch B of the bifurcation.\newline
\cite{Belokurov2006} states that the SGB of branch B is significantly brighter and hence probably slightly closer than A, but the branch itself is reported to have much lower luminosity compared to A.

\begin{figure*}
\begin{center}  
\includegraphics[]{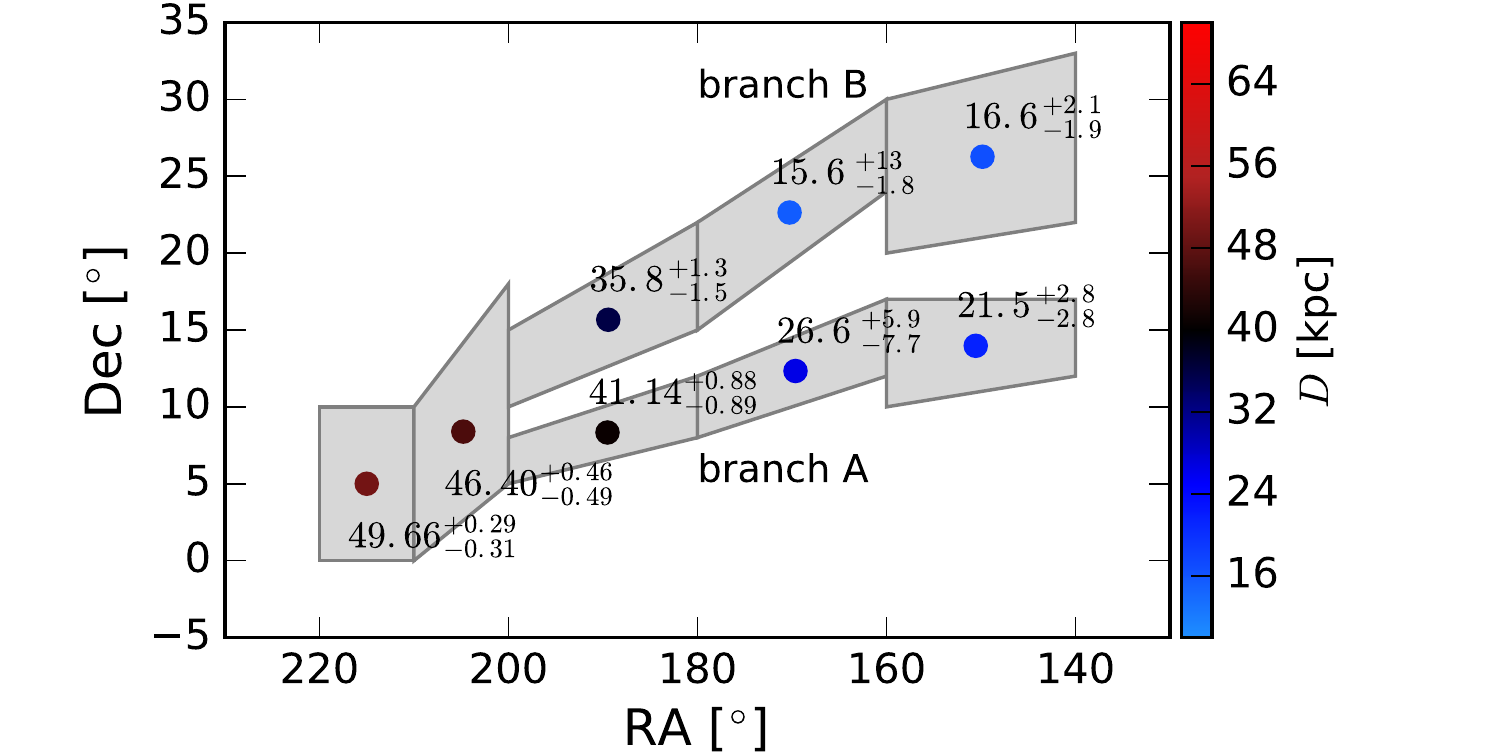}
\caption[Heliocentric distance estimates for patches covering the branches A and B at lower and upper declination, respectively \citep{Belokurov2006}, of the bifurcated Sagittarius stream.]{{Heliocentric distance estimates for patches covering the branches A and B of the Sagittarius stream in equatorial coordinates.
For each patch, the fit using the halo and stream model as described above in Section \ref{sec:TheModelSgr} was carried out derive distance estimates. The points set at the centroid of each polygon indicate the heliocentric distance $D$ in kpc as estimated from the sample within each polygon. The $D_{\mathrm{sgr}}\pm \delta D_{\mathrm{sgr}}$ range is indicated.}
\label{fig:sgr_radec_fitted_distances_larger_polygon}}
\end{center}  
\end{figure*}

Their Fig. 4 shows a noticeable, but small difference in the distances estimated for branches A and B of 3 to $15$~kpc, qualitatively consistent with the simulations by \cite{Fellhauer2006}. However, \cite{Ruhland2011} found from an analysis of BHB stars in the stream that the branches differ by at most 2~kpc in distance.
To follow up on this, we measured the RRab mean distances  for small patches in both branches, as shown by the polygons in Fig. \ref{fig:sgr_radec_fitted_distances_larger_polygon}, fitting a halo and stream model as described above in Section \ref{sec:TheModelSgr}.
This fitting led to the distance estimates as shown in Fig. \ref{fig:sgr_radec_fitted_distances_larger_polygon} and in Tab. \ref{tab:sgr_bifurcation} in the Table Appendix.
Indeed a small distance difference between the two branches can be found, branch B being closer than branch A like in the simulation by \cite{Fellhauer2006}. But the sparse sampling by the RR  Lyrae makes this analysis inconclusive. 

\subsection{Bifurcation of the Trailing Arm}
\label{sec:TrailingBifurcation}

Analogous to the bifurcation of the leading arm found by \cite{Belokurov2006}, \cite{Koposov2012} found a similar bifurcation in the Sgr stream trailing arm,
consisting of two branches that are separated on the sky by ${\sim}10\arcdeg$.

These bifurcation was later confirmed and studied in greater detail by \cite{Slater2013}, using main-sequence turn-off (MSTO) and red clump (RC) stars from the Pan-STARRS1 survey, and \cite{Navarrete2017}, who have examined a large portion of approximately $65\arcdeg$ of the Sgr trailing arm available in the imaging data from the VST ATLAS survey, using BHB and SGB stars, as well as RR Lyrae from CRTS.

They found the trailing arm appearing to be split along the line-of-sight, with the additional stream component following a distinct distance track, and a difference in heliocentric distances exists of ${\sim}5$ kpc. The bulk of the ``bright stream'' \citep{Slater2013} is below the Sgr orbital plane (thus $\tilde{B}_{\odot}<0\arcdeg$), while the ``faint stream'' lies mostly above the plane ( $\tilde{B}_{\odot}>0\arcdeg$).

We compare here our distance distributions to the findings of \cite{Slater2013} and \cite{Navarrete2017} for different regions in $( \tilde{\Lambda}_{\odot}, \tilde{B}_{\odot})$.

\cite{Navarrete2017} report a bifurcation in the $( \tilde{\Lambda}_{\odot}, \tilde{B}_{\odot})$ plane with a separation of ${\sim}10 \arcdeg$.
Likely due to our relatively sparse source density, we can not find an indicator for a bifurcation in the $( \tilde{\Lambda}_{\odot}, \tilde{B}_{\odot})$ plane that would lead to a ``bright stream'' and ``faint stream''.

We then checked whether we can identify l.o.s. substructures, and made histograms of the heliocentric distance distribution for several patches along the trailing arm of the Sgr stream.

\begin{figure*}
\begin{center}  
\includegraphics[trim=0.0inch 0.0inch 0.0inch 0.15inch, clip=true]{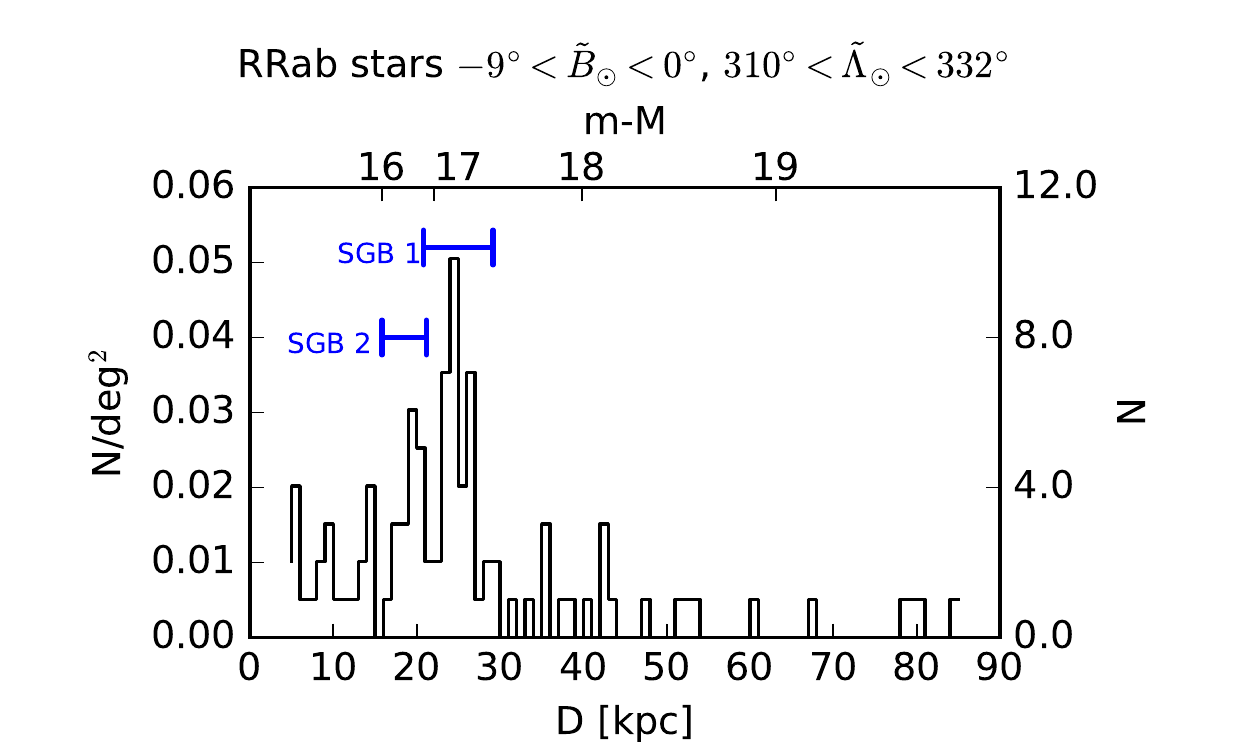}
\caption[Heliocentric distance distribution for RRab stars in the trailing tail, describing bifurcation.]{{Heliocentric distance distribution for RRab stars in the trailing tail. Analogous to the bifurcation of the leading arm found by \cite{Belokurov2006}, \cite{Slater2013} and \cite{Navarrete2017} report a similar bifurcation in the trailing tail. We compare our distance distributions to the findings of \cite{Slater2013} and \cite{Navarrete2017} for different regions in $( {\tilde{B}_{\odot}}, \tilde{\Lambda}_{\odot})$. This plot gives our distance estimates in a region also probed using RR Lyrae by \cite{Navarrete2017} (see their Fig. 9). We give our estimates of the heliocentric distance $D$ and the distance modulus $m-M$ \citep{Sesar2017b}. Blue markers represent substructures found by \cite{Navarrete2017}.}
\label{fig:bifurcation_trailing}}
\end{center}  
\end{figure*}

In Fig. \ref{fig:bifurcation_trailing}, we give a histogram of our distance estimates in one of the regions probed by \cite{Navarrete2017} and \cite{Slater2013}. This specific region was also probed using RR Lyrae by \cite{Navarrete2017} (see their Fig. 9). We give our estimates of the heliocentric distance $D$ and the distance modulus $m-M$ \citep{Sesar2017b}. Blue markers represent substructures found by \cite{Navarrete2017}. A similar shape of the distance distribution is found, and we also detect the substructures they call ``SGB 1'' and ``SGB 2''. We find ``SGB 1'' at a slightly larger distance than \cite{Navarrete2017}. We find ``SGB 2'' split into two components.

We were also able to identify similar substructures as found by \cite{Navarrete2017} and \cite{Slater2013} within other patches of the Sgr stream trailing arm, and count them as tentative but marginal significant because the relatively low density of our tracers.

\section{Summary}

In this work, we quantified the geometry of the Sagittarius stream, characterizing the l.o.s. density of the Sagittarius stream approximated by a Gaussian distribution centered on the distance $D_{\mathrm{sgr}}$, having the l.o.s. depth $\sigma_{\mathrm{sgr}}$. This model was used to estimate distance and depth of the Sgr stream as given by RR Lyrae candidates (RRab with completeness$\geq$ 0.8, purity=0.9 up to 80~kpc, distance precision of $3\%$) resulting from the classification that incorporates period fitting.

The fitting resulted into the best and first basically complete (i.e., spanning $0\arcdeg<\tilde{\Lambda}_{\odot}<360\arcdeg$) trace of Sgr stream's heliocentric distance, as well as l.o.s. depth. This model allows further to measure many properties of the Sgr stream. We have measured the depth $\sigma_{\mathrm{sgr}}$ as well as the deprojected depth of the stream. The function of $\sigma_{\mathrm{sgr}}$ vs. $\tilde{\Lambda}_{\odot}$ can be partially explained by projection effects, and partially by projection effects due to the angle our line-of-sight direction forms with the stream direction.
Deprojection removes the line-of-sight effects and thus results into a depth of the stream that will be very helpful when comparing simulations to observational data.
Further on, we computed the amplitude of the Str stream as the number of RRab stars in the sream per degree as a function of its longitude $\tilde{\Lambda}_{\odot}$.
The fit allows us to precisely determine the apocenter positions, from which we then calculate the orbital precession. We also find a strong indicator for a precession of the orbital plane.
We have measured the Galactocentric angle between the apocenters of the leading and trailing arm of the Sgr stream and the difference between their respective distances.\\
Having now a model of the geometry of the Sgr stream at hand, it can be used to further constrain the Milky Way's potential.

\acknowledgments

N.H., B.S.and H.-W.R. acknowledge funding from the
European Research Council under the European Unions
Seventh Framework Programme (FP 7) ERC Grant Agreement n. [321035].

The Pan-STARRS1 Surveys (PS1) have been made possible through contributions of the Institute for Astronomy, the University of Hawaii, the Pan-STARRS Project Office, the Max-Planck Society and its participating institutes, the Max Planck Institute for Astronomy, Heidelberg and the Max Planck Institute for Extraterrestrial Physics, Garching, The Johns Hopkins University, Durham University, the University of Edinburgh, Queen's University Belfast, the Harvard-Smithsonian Center for Astrophysics, the Las Cumbres Observatory Global Telescope Network Incorporated, the National Central University of Taiwan, the Space Telescope Science Institute, the National Aeronautics and Space Administration under Grant No. NNX08AR22G issued through the Planetary Science Division of the NASA Science Mission Directorate, the National Science Foundation under Grant No. AST-1238877, the University of Maryland, and Eotvos Lorand University (ELTE) and the Los Alamos National Laboratory. 

\clearpage

\appendix

\section{Tables}
\label{sec:Tables}

Table \ref{tab:table_RRLyr} gives the PS1 RRab stars with $\vert \tilde{B}_{\odot} \vert <-9\arcdeg$ this analysis is based on.

Tables \ref{tab:Dprior_leading} and \ref{tab:Dprior_trailing} give the minimum and maximum prior on $D_{\mathrm{sgr}}$, $D_{\mathrm{minprior}}$ and $D_{\mathrm{maxprior}}$, as indicated in Fig. \ref{fig:bestfit}. The annotation ``max'' within the tables state that the given value is the maximum observed heliocentric distance $D$ in the given $\tilde{\Lambda}_{\odot}$ interval. 

Tables \ref{tab:sgr_leading_arm} and \ref{tab:sgr_trailing_arm} give the geometry of the Sagittarius stream, represented by its extent and depth as inferred from the analysis presented in this paper.

Tables \ref{tab:deprojected_leading} and \ref{tab:deprojected_trailing} give the deprojected depth of the Sagittarius stream.

Tables \ref{tab:amplitude_weighted_B_leading} and \ref{tab:amplitude_weighted_B_trailing} give the amplitude of the Sagittarius stream. as well as the weighted latitude $\langle{\tilde{B}_{\odot}}\rangle$ of the stream, as calculated by Equ. \eqref{equ:weighted_B_o}.

Table \ref{tab:sgr_bifurcation} gives the distance estimates for the branches A and B of the Sagittarius stream.

\capstartfalse
\begin{deluxetable}{ccccccc}
\tabletypesize{\scriptsize}
\tablecolumns{7}
\tablecaption{PS1 RRab Stars with $\vert \tilde{B}_{\odot} \vert <-9\arcdeg$\label{tab:table_RRLyr}}
\tablehead{
\colhead{RA} & \colhead{Dec} & \colhead{$score_{\rm 3,ab}^a$} & \colhead{DM$^b$} & \colhead{Period} & \colhead{$\phi_0^c$} & \colhead{$A_r^d$} \\
\colhead{(deg)} & \colhead{(deg)} & \colhead{$ $} & \colhead{(mag)} & \colhead{(day)} & \colhead{(day)} & \colhead{(mag)}
}
\startdata
181.40332 & 7.77677 & 1.00 & 17.08 & 0.6752982619 &  0.24526 & 0.75 \\
181.12043 & 8.28025 & 0.91 & 12.79 & 0.5382632283 &  0.44801 & 0.63 \\
180.08748 & 9.10501 & 1.00 & 17.76 & 0.5203340425 & -0.46188 & 0.88 \\
\enddata
\tablenotetext{a}{Final RRab classification score.}
\tablenotetext{b}{Distance modulus. The uncertainty in distance modulus is $0.06(rnd)\pm0.03(sys)$ mag.}
\tablenotetext{c}{Phase offset (see Equation 2 of \citealt{Sesar2017b}).}
\tablenotetext{d}{PS1 $r$-band light curve amplitude.}
\tablecomments{A machine readable version of this table is available in the electronic edition of the Journal. A portion is shown here for guidance regarding its form and content.} 
\end{deluxetable}
\capstarttrue

\capstartfalse
\begin{deluxetable}{ccc}
\tabletypesize{\scriptsize}
\tablecolumns{3}
\tablecaption{$D_{\mathrm{sgr}}$ Prior, Leading Arm\label{tab:Dprior_leading}}
\tablehead{
\colhead{$\tilde{\Lambda}_{\odot}$ interval (deg)}  &  \colhead{$D_{\mathrm{minprior}}$ (kpc)}  & \colhead{$D_{\mathrm{maxprior}}$ (kpc)}}
\startdata
$[10,20[$  &    5 & 27.3 (max)\\      
$[20,30[$  &    30 & 35\\      
$[30,40[$  &    30 & 37\\
$[40,50[$  &    10 & 77.5 (max)\\
$[50,60[$  &    10 & 110.1 (max)\\
$[60,70[$  &    10 & 98.9 (max)\\
$[70,80[$  &    10 & 97.2 (max)\\  
$[80,90[$ &     20 & 70\\      
$[90,100[$	&  20 & 60\\	  
$[100,110[$ &	 25 & 50 \\
$[110,120[$	 & 20 & 50\\
$[120,130[$ &	 15 & 40\\	  
$[130,140[$	&  20 & 40\\	  
$[140,150[$	&  15 & 40\\
$[150,160[$	& 15 & 40\\
\enddata
\end{deluxetable}   
\capstarttrue

\capstartfalse
\begin{deluxetable}{ccc}
\tabletypesize{\scriptsize}
\tablecolumns{3}
\tablecaption{$D_{\mathrm{sgr}}$ Prior, Trailing Arm\label{tab:Dprior_trailing}}
\tablehead{
\colhead{$\tilde{\Lambda}_{\odot}$ interval (deg)}  &  \colhead{$D_{\mathrm{minprior}}$ (kpc)}  & \colhead{$D_{\mathrm{maxprior}}$ (kpc)}}
\startdata
$[100,110[$ & 50 & 95.1 (max)\\
$[110,120[$ & 50 & 98.9 (max)\\
$[120,130[$ & 40 & 92.6 (max)\\
$[130,140[$ & 10 & 92.7 (max)\\
$[140,150[$ & 40 & 106.0 (max)\\
$[150,160[$ & 40 & 134.2 (max)\\
$[160,170[$ & 10 & 125.1 (max)\\
$[170,180[$ & 10 & 131.7 (max)\\
$[180,190[$ & 10 & 103.6 (max)\\
$[190;200[$ & 10 & 72.6 (max)\\
$[200,210[$ & 40 & 73.8 (max)\\
$[210,220[$ & 10 & 64.6 (max)\\
$[220,230[$ & 30 & 60\\
$[230,240[$ & 30 & 60\\
$[240,250[$ & 20 & 50\\
$[250,260[$ & 10 & 40\\
$[260,270[$ & 10 & 50\\
$[270,280[$ & 10 & 50\\
$[280,290[$ & 10 & 50\\
$[290,300[$ & 10 & 50\\
$[300,310[$ & 10 & 50\\
$[310,320[$ & 10 & 80.0 (max)\\
$[320,330[$ & 10 & 84.7 (max)\\
$[330,340[$ & 10 &  64.6 (max)\\
$[340,350[$ & 10 & 86.4 (max)\\
$[350,360[$ & 10 & 50.0\\
\enddata
\end{deluxetable}   
\capstarttrue

\clearpage

\capstartfalse
\begin{deluxetable}{cccccccccccc}
\tabletypesize{\scriptsize}
\tablecolumns{12}
\tablecaption{Fitted Parameters for Sagittarius Stream, Leading Arm\label{tab:sgr_leading_arm}}
\tablehead{
\colhead{$\tilde{\Lambda}_{\odot}$ (deg)}  & \colhead{$f_{\mathrm{sgr}}$ $^a$} &  \colhead{$D_{\mathrm{sgr}}$ (kpc)$^b$} & \colhead{$\delta_{-}(D_{\mathrm{sgr}})$} &  \colhead{$\delta_{+}(D_{\mathrm{sgr}})$} & \colhead{$2\delta_{-}(D_{\mathrm{sgr}}) $} & \colhead{$2\delta_{+}(D_{\mathrm{sgr}})$} & \colhead{$\sigma_{\mathrm{sgr}}$ (kpc)$^c$} & \colhead{$\delta_{-}(\sigma_{\mathrm{sgr}})$} & \colhead{$\delta_{+}(D_{\mathrm{sgr}})$} & \colhead{$2\delta_{-}(\sigma_{\mathrm{sgr}})$} & \colhead{$2\delta_{+}(\sigma_{\mathrm{sgr}})$}}
\startdata
5 & 0.18 & 28.830 & 0.10 & 0.094 & 0.20 & 0.18 & 1.621 & 0.079 & 0.091 & 0.15 & 0.18\\
15 & 0.052 & 14.3 & 7.1 & 9.7 & 9.1 & 12.0 & 2.8 & 1.5 & 6.5 & 1.7 & 15\\
25 & 0.050 & 34.14 & 1.8 & 0.66 & 3.8 & 0.83 & 2.8 & 1.4 & 3.3 & 1.7 & 9.3\\
35 & 0.051 & 36.65 & 0.60 & 0.27 & 1.9 & 0.34 & 4.1 & 1.7 & 1.9 & 2.9 & 4.6\\
45 & 0.38 & 45.94 & 0.24 & 0.25 & 0.48 & 0.52 & 3.68 & 0.19 & 0.20 & 0.36 & 0.43\\
55 & 0.51 & 50.50 & 0.17 & 0.17 & 0.34 & 0.33 & 3.33 & 0.18 & 0.18 & 0.34 & 0.38\\
65 & 0.61 & 52.59 & 0.21 & 0.21 & 0.43 & 0.44 & 4.52 & 0.27 & 0.27 & 0.54 & 0.53\\
75 & 0.41 & 49.19 & 0.26 & 0.27 & 0.52 & 0.53 & 3.75 & 0.30 & 0.33 & 0.57 & 0.72\\
85 & 0.36 & 46.22 & 0.40 & 0.39 & 0.83 & 0.78 & 4.66 & 0.44 & 0.47 & 0.79 & 0.99\\
95 & 0.21 & 40.59 & 0.48 & 0.53 & 0.95 & 1.1 & 3.88 & 0.72 & 0.83 & 1.3 & 1.9\\
105 & 0.26 & 34.8 & 6.7 & 1.9 & 9.4 & 2.8 & 6.3 & 2.2 & 6.0 & 3.2 & 8.8\\
115 & 0.22 & 31.19 & 0.57 & 0.54 & 1.3 & 1.0 & 3.08 & 0.51 & 0.66 & 0.91 & 1.7\\
125 & 0.25 & 25.9 & 5.9 & 1.9 & 9.9 & 3.0 & 5.0 & 1.9 & 3.7 & 2.8 & 6.2\\
135 & 0.067 & 21.34 & 0.92 & 2.1 & 1.3 & 8.9 & 2.7 & 1.3 & 2.7 & 1.7 & 6.5\\
145 & 0.13 & 19.66 & 0.87 & 0.80 & 2.2 & 20.0 & 2.05 & 0.68 & 1.0 & 0.99 & 2.5\\
155 & 0.15 & 16.2 & 1.1 & 3.2 & 1.1 & 3.2 & 3.4 & 2.2 & 2.6 & 2.2 & 2.6\\
\enddata
\tablenotetext{a}{fraction sources in Sgr stream}
\tablenotetext{b}{mean heliocentric Sgr stream distance}
\tablenotetext{c}{Sgr stream line-of-sight depth} 
\end{deluxetable}   
\capstarttrue

\capstartfalse
\begin{deluxetable}{cccccccccccc}
\tablecolumns{12}
\tabletypesize{\scriptsize}
\tablecaption{Fitted Parameters for Sagittarius Stream, Trailing Arm\label{tab:sgr_trailing_arm}}
\tablehead{
\colhead{$\tilde{\Lambda}_{\odot}$ (deg)}  & \colhead{$f_{\mathrm{sgr}}$ $^a$} &  \colhead{$D_{\mathrm{sgr}}$ (kpc)$^b$} & \colhead{$\delta_{-}(D_{\mathrm{sgr}})$} &  \colhead{$\delta_{+}(D_{\mathrm{sgr}})$} & \colhead{$2\delta_{-}(D_{\mathrm{sgr}}) $} & \colhead{$2\delta_{+}(D_{\mathrm{sgr}})$} & \colhead{$\sigma_{\mathrm{sgr}}$ (kpc)$^c$} & \colhead{$\delta_{-}(\sigma_{\mathrm{sgr}})$} & \colhead{$\delta_{+}(D_{\mathrm{sgr}})$} & \colhead{$2\delta_{-}(\sigma_{\mathrm{sgr}})$} & \colhead{$2\delta_{+}(\sigma_{\mathrm{sgr}})$}}
\startdata
105 & 0.055 & 55.4 & 3.9 & 1.9 & 5.2 & 3.5 & 3.2 & 1.5 & 7.2 & 2.0 & 13\\
115 & 0.056 & 62.3 & 6.0 & 5.8 & 11 & 10 & 3.5 & 2.2 & 7.3 & 2.5 & 14\\
125 & 0.059 & 57.2 & 1.9 & 1.4 & 11 & 11 & 2.3 & 0.97 & 2.2 & 1.3 & 12\\
135 & 0.084 & 66.9 & 2.6 & 3.2 & 5.0 & 7.2 & 5.8 & 2.4 & 4.0 & 4.1 & 9.9\\
145 & 0.095 & 81.3 & 4.1 & 2.7 & 8.7 & 4.4 & 6.1 & 3.1 & 3.0 & 4.6 & 6.2\\
155 & 0.31 & 83.1 & 1.9 & 1.8 & 1.9 & 1.8 & 5.2 & 1.7 & 3.0 & 1.7 & 2.0\\
165 & 0.36 & 89.02 & 0.72 & 0.74 & 1.5 & 1.5 & 5.13 & 0.64 & 0.85 & 1.2 & 2.0\\
175 & 0.63 & 92.98 & 0.79 & 0.81 & 1.6 & 1.6 & 8.99 & 0.64 & 0.68 & 1.3 & 1.4\\
185 & 0.40 & 86.7 & 3.0 & 2.2 & 7.0 & 4.6 & 10.5 & 2.8 & 5.2 & 4.6 & 8.7\\
195 & 0.082 & 60.0 & 7.1 & 9.0 & 17 & 12 & 2.8 & 1.5 & 5.8 & 1.8 & 15\\
205 & 0.55 & 53.0 & 1.2 & 1.1 & 2.6 & 2.1 & 6.78 & 0.82 & 0.95 & 1.5 & 2.2\\
215 & 0.61 & 43.15 & 1.1 & 0.88 & 2.4 & 1.8 & 6.65 & 0.97 & 1.2 & 1.8 & 2.4\\
225 & 0.71 & 36.55 & 0.87 & 0.75 & 1.8 & 1.4 & 6.28 & 0.49 & 0.61 & 0.97 & 1.4\\
235 & 0.55 & 31.17 & 0.70 & 0.80 & 1.1 & 1.6 & 6.16 & 0.65 & 0.71 & 1.3 & 1.5\\
245 & 0.58 & 28.41 & 0.85 & 0.69 & 1.9 & 1.3 & 4.66 & 0.60 & 0.75 & 1.1 & 1.7\\
255 & 0.62 & 25.57 & 0.85 & 0.72 & 1.9 & 1.4 & 5.14 & 0.54 & 0.64 & 1.0 & 1.4\\
265 & 0.43 & 24.7 & 1.2 & 0.87 & 2.8 & 1.6 & 4.86 & 0.90 & 1.1 & 1.7 & 2.4\\
275 & 0.60 & 18.0 & 3.1 & 2.1 & 7.0 & 3.9 & 7.7 & 1.8 & 2.1 & 3.3 & 4.2\\
285 & 0.32 & 20.34 & 1.1 & 0.83 & 2.7 & 1.6 & 4.44 & 0.67 & 0.90 & 1.2 & 2.2\\
295 & 0.27 & 21.2 & 1.8 & 1.2 & 5.4 & 2.2 & 4.7 & 1.2 & 1.7 & 2.0 & 4.2\\
305 & 0.37 & 20.8 & 1.3 & 1.0 & 3.4 & 1.9 & 5.17 & 0.88 & 1.1 & 1.6 & 2.8\\
315 & 0.45 & 21.66 & 0.95 & 0.80 & 2.0 & 1.5 & 4.84 & 0.67 & 0.80 & 1.3 & 1.8\\
325 & 0.48 & 22.00 & 0.75 & 0.62 & 1.7 & 1.2 & 4.41 & 0.52 & 0.63 & 0.97 & 1.5\\
335 & 0.40 & 20.1 & 2.3 & 1.4 & 7.3 & 2.3 & 5.3 & 1.2 & 1.9 & 2.1 & 4.6\\
345 & 0.47 & 19.7 & 1.7 & 1.3 & 4.7 & 2.4 & 6.43 & 0.73 & 0.92 & 1.4 & 2.3\\
355 & 0.43 & 27.605 & 0.054 & 0.053 & 0.11 & 0.11 & 1.245 & 0.048 & 0.047 & 0.091 & 0.096\\
\enddata
\tablenotetext{a}{fraction sources in Sgr stream}
\tablenotetext{b}{mean heliocentric Sgr stream distance}
\tablenotetext{c}{Sgr stream line-of-sight depth} 
\end{deluxetable}   
\capstarttrue

\clearpage

\capstartfalse
\begin{deluxetable}{cccc}
\tabletypesize{\scriptsize}
\tablecolumns{12}
\tablecaption{Deprojected Depth $\tilde{\sigma}_{\mathrm{sgr}}$ for Sagittarius Stream (see Sec. \ref{sec:TheDepthOfTheSagittariusStream}), Leading Arm\label{tab:deprojected_leading}}
\tablehead{
\colhead{$\tilde{\Lambda}_{\odot}$ (deg)}  & \colhead{$\tilde{\sigma}_{\mathrm{sgr}}$} &  \colhead{$\delta_{-}(\tilde{\sigma}_{\mathrm{sgr}})$} & \colhead{$\delta_{+}\tilde{\sigma}_{\mathrm{sgr}}$}}
\startdata
15 & 1.79 & 0.83 & 6.0\\
25 & 2.6 & 1.3 & 5.7\\
35 & 3.6 & 2.1 & 5.2\\
45 & 2.7 & 2.6 & 2.9\\
55 & 3.1 & 2.9 & 3.3\\
65 & 4.5 & 4.2 & 4.8\\
75 & 3.5 & 3.2 & 3.8\\
85 & 4.1 & 3.7 & 4.5\\
95 & 3.0 & 2.5 & 3.7\\
105 & 5.1 & 3.3 & 9.9\\
115 & 2.4 & 2.0 & 2.9\\
125 & 3.4 & 2.2 & 6.0\\
135 & 2.2 & 1.2 & 4.3\\
145 & 1.7 & 1.2 & 2.6\\
\enddata
\end{deluxetable}   
\capstarttrue

\capstartfalse
\begin{deluxetable}{cccc}
\tablecolumns{12}
\tabletypesize{\scriptsize}
\tablecaption{Deprojected Depth $\tilde{\sigma}_{\mathrm{sgr}}$ for Sagittarius Stream (see Sec. \ref{sec:TheDepthOfTheSagittariusStream}), Trailing Arm\label{tab:deprojected_trailing}}
\tablehead{
\colhead{$\tilde{\Lambda}_{\odot}$ (deg)}  & \colhead{$\tilde{\sigma}_{\mathrm{sgr}}$} &  \colhead{$\delta_{-}(\tilde{\sigma}_{\mathrm{sgr}})$} & \colhead{$\delta_{+}\tilde{\sigma}_{\mathrm{sgr}}$}}
\startdata
115 & 0.61 & 0.24 & 1.9\\
125 & 2.2 & 1.3 & 4.4\\
135 & 4.2 & 2.5 & 7.0\\
145 & 5.2 & 2.5 & 7.7\\
155 & 5.1 & 3.4 & 7.0\\
165 & 4.9 & 4.3 & 5.7\\
175 & 9.0 & 8.3 & 9.6\\
185 & 6.6 & 4.8 & 9.9\\
195 & 1.65 & 0.76 & 5.1\\
205 & 5.0 & 4.4 & 5.7\\
215 & 4.6 & 3.9 & 5.4\\
225 & 4.6 & 4.3 & 5.1\\
235 & 5.0 & 4.5 & 5.6\\
245 & 4.1 & 3.5 & 4.7\\
255 & 4.8 & 4.3 & 5.4\\
265 & 3.5 & 2.8 & 4.3\\
275 & 6.8 & 5.2 & 8.6\\
285 & 4.0 & 3.4 & 4.9\\
295 & 4.7 & 3.5 & 6.3\\
305 & 5.2 & 4.3 & 6.3\\
315 & 4.8 & 4.1 & 5.6\\
325 & 4.3 & 3.8 & 4.9\\
335 & 5.1 & 3.9 & 6.9\\
345 & 4.1 & 3.7 & 4.7\\
\enddata
\end{deluxetable}   
\capstarttrue

\clearpage

\capstartfalse
\begin{deluxetable}{ccc}
\tablecolumns{12}
\tabletypesize{\scriptsize}
\tablecaption{Amplitude $A$ (see Sec. \ref{sec:TheAmplitudeOfTheSagittariusStream}) and weighted latitude $\langle{\tilde{B}_{\odot}}\rangle$ (see Sec. \ref{sec:TheOrbitalPlanePrecessionoftheSagittariusStream}) for Sagittarius Stream, Leading Arm\label{tab:amplitude_weighted_B_leading}}
\tablehead{
\colhead{$\tilde{\Lambda}_{\odot}$ (deg)}  & \colhead{$A$ (deg$^{-1} \times$ kpc$^{-1}$)} & \colhead{$\langle{\tilde{B}_{\odot}}\rangle$ (deg)}}
\startdata
5 & 24 & 0.48\\
15 & 0.53 & 4.3\\
25 & 3.9 & -0.41\\
35 & 1.2 & 0.40\\
45 & 7.4 & 0.90\\
55 & 11 & 0.14\\
65 & 9.5 & -0.11\\
75 & 5.4 & 0.39\\
85 & 3.0 & -0.67\\
95 & 1.9 & -0.47\\
105 & 1.1 & -0.74\\
115 & 1.5 &  -1.4\\
125 & 0.70 &  -0.65\\
135 & 0.44 &  -1.4\\
145 & 1.0 &  -0.25\\
155 & 0.62 &  -2.0\\
\enddata
\end{deluxetable}   
\capstarttrue

\clearpage

\capstartfalse
\begin{deluxetable}{ccc}
\tablecolumns{12}
\tabletypesize{\scriptsize}
\tablecaption{Amplitude $A$ (see Sec. \ref{sec:TheAmplitudeOfTheSagittariusStream}) and weighted latitude $\langle{\tilde{B}_{\odot}}\rangle$ (see Sec. \ref{sec:TheOrbitalPlanePrecessionoftheSagittariusStream}) for Sagittarius Stream, Leading Arm\label{tab:amplitude_weighted_B_trailing}}
\tablehead{
\colhead{$\tilde{\Lambda}_{\odot}$ (deg)}  & \colhead{$A$ (deg$^{-1} \times$ kpc$^{-1}$)} & \colhead{$\langle{\tilde{B}_{\odot}}\rangle$ (deg)}}
\startdata
105 & 0.47 & -2.1\\
115 & 0.33 & -0.51\\
125 & 0.37 & -0.96\\
135 & 0.26 & -1.5\\
145 & 0.25 & -2.0\\
155 & 0.85 & -1.8\\
165 & 1.2 & -1.2\\
175 & 1.6 & 0.59\\
185 & 0.46 & -0.18\\
195 & 0.067 & -4.2\\
205 & 0.93 & -2.6\\
215 & 1.2 & -0.60\\
225 & 1.8 & -0.15\\
235 & 1.2 & -0.34\\
245 & 1.6 & 0.91\\
255 & 1.6 & 0.55\\
265 & 1.2 & -0.12\\
275 & 1.1 & -0.48\\
285 & 1.1 & -0.08\\
295 & 1.3 & -0.20\\
305 & 1.6 & 1.3\\
315 & 1.8 & 2.4\\
325 & 3.1 & 4.2\\
335 & 2.3 & 4.3\\
345 & 3.6 & 4.3\\
355 & 54 & 0.88\\
\enddata
\end{deluxetable}   
\capstarttrue

\clearpage

\capstartfalse
\begin{deluxetable}{ccccccccccccc}
\tablecolumns{14}
\tabletypesize{\scriptsize}
\tablecaption{Possibly Sagittarius Stream Bifurcation\label{tab:sgr_bifurcation}}
\tablewidth{0pt}
\tablehead{
\colhead{RA (deg) $^a$}  & \colhead{Dec (deg) $^a$}  & \colhead{$f_{\mathrm{sgr}}$ $^b$} &  \colhead{$D_{\mathrm{sgr}}$ (kpc)$^c$} 
& \colhead{$\delta_{-}(D_{\mathrm{sgr}})$} &  \colhead{$\delta_{+}(D_{\mathrm{sgr}})$} & \colhead{$2\delta_{-}(D_{\mathrm{sgr}}) $} & \colhead{$2\delta_{+}(D_{\mathrm{sgr}})$} & \colhead{$\sigma_{\mathrm{sgr}}$ (kpc)$^c$} & \colhead{$\sigma_{-}(\sigma_{\mathrm{sgr}})$} & \colhead{$\sigma_{+}(D_{\mathrm{sgr}})$} & \colhead{$2\sigma_{-}(\sigma_{\mathrm{sgr}})$} & \colhead{$2\sigma_{+}(\sigma_{\mathrm{sgr}})$}}
\startdata
215 & 5 & 0.49 & 49.77 & 0.31 & 0.32 & 0.63 & 0.66 & 3.18 & 0.43 & 0.52 & 0.83 & 1.1\\
204.783 & 8.391 & 0.41 & 46.37 & 0.51 & 0.49 & 1.0 & 0.94 & 4.56 & 0.52 & 0.57 & 1.0 & 1.2\\
189.524 & 8.333 & 0.32 & 41.22 & 1.1 & 0.87 & 2.8 & 1.7 & 5.48 & 1.0 & 1.3 & 1.9 & 3.3\\
189.444 & 15.667 & 0.44 & 20.0 & 6.0 & 6.6 & 9.1 & 15 & 16.14 & 3.5 & 2.5 & 8.9 & 3.6\\
169.63 & 12.333 & 0.40 & 21.7 & 7.8 & 6.5 & 11 & 12 & 12.0 & 4.2 & 4.3 & 8.7 & 6.8\\
170.256 & 22.641 & 0.33 & 17 & 5.8 & 12 & 7.4 & 13 & 10.1 & 7.4 & 2.6 & 8.7 & 4.5\\
150.556 & 13.972 & 0.15 & 25.18 & 1.4 & 0.96 & 4.5 & 1.7 & 2.0 & 0.72 & 1.4 & 0.99 & 4.2\\
149.841 & 26.27 & 0.11 & 19.46 & 4.3 & 1.4 & 8.7 & 50 & 2.9 & 1.5 & 4.0 & 1.9 & 10\\
\enddata
\tablenotetext{a}{for each polygon, the centroid of its $(\alpha,\delta)$ is given, as used in Fig. \ref{fig:sgr_radec_fitted_distances_larger_polygon}.}
\tablenotetext{b}{fraction sources in Sgr stream}
\tablenotetext{c}{mean heliocentric Sgr stream distance}
\tablenotetext{d}{Sgr stream line-of-sight depth} 
\end{deluxetable}   
\capstarttrue


\clearpage

\begin{thebibliography}{}
\bibitem[Belokurov et al. (2006)]{Belokurov2006} Belokurov, V., Zucker, D.~B., Evans, N.~W., et al. 2006, \apj, 642, 2, L137
\bibitem[Belokurov et al. (2014)]{Belokurov2014} Belokurov, V., Koposov, S.~E., Evans, N.~W., 2014, \mnras, 437, 1
\bibitem[Binney et al. (2008)]{Binney2008} Binney J., Tremaine S., 2008, Galactic Dynamics, 2nd edn. Princeton Univ. Press, Princeton, NJ
\bibitem[Bovy et al. (2016)]{Bovy2016} Bovy, J., Bahmanyar, A., Fritz, T.~K., et al. 2016, \aj, 883, 31
\bibitem[Chambers et al. (2016)]{Chambers2016} Chambers, K.~C., Magnier, E.~A., Metcalfe, N., et al. 2016, arXiv:1612.05560 [astro-ph.IM]
\bibitem[Dierickx \& Loeb (2017)]{Dierickx2017} Dierickx, M.~I.~P., Loeb, A., 2017, \apj, 836, 1, 92
\bibitem[Drake et al. (2013a)]{Drake2013a} Drake, A.~J., Catelan, M., Djorgovski, S.~G., et al. 2013, \apj, 765, 154
\bibitem[Drake et al. (2014)]{Drake2014} Drake, A.~J., Graham, M.~J, Djorgovski, S.~G., et al. 2014, \apjs, 213, 1, 9	
\bibitem[Duffau et al. (2014)]{Duffau2014} Duffau, S., Vivas, A.~K., Zinn, R., et al. 2014, \ap, 566, A118
\bibitem[Eyre \& Binney (2009)]{Eyre2009} Eyre, A., \& Binney, J. 2009, \mnras, 400, 548
\bibitem[Fardal et al. (2015)]{Fardal2015} Fardal, M.~A., Huang, S., Weinberg, M.~D., 2015, \mnras, 452, 301 
\bibitem[Fellhauer et al. (2006)]{Fellhauer2006} Fellhauer, M., Belokurov, V., Evans, N.~W., et al. 2006, \apj, 651, 1, 167
\bibitem[Foreman et al. (2012)]{Foreman2012} Foreman-Mackey, D., Hogg, D.~W., Lang, D., et al. 2012, arXiv:1202.3665 [astro-ph.IM]]
\bibitem[Gaia Collaboration et al.(2016)]{Gaia2016} Gaia Collaboration, Prusti, T., de Bruijne, J.~H.~J., et al.\ 2016, \aap, 595, A1
\bibitem[Gibbons et al. (2014)]{Gibbons2014} Gibbons, S.~L.~J., Belokurov, V., Evans, N.~W. 2014, \mnras, 445, 4, 3788
\bibitem[Goodman \& Weare (2010)]{Goodman2010} Goodman, J., Weare, J., 2010, Communications in Applied Mathematics and Computational Science, 5
\bibitem[Helmi (2004a)]{Helmi2004a} Helmi, A. 2004a, \mnras, 351, 643
\bibitem[Helmi (2004b)]{Helmi2004b} Helmi, A. 2004b, ApJL, 610, L97
\bibitem[Hernitschek et al. (2016)]{Hernitschek2016} Hernitschek, N., Rix., H.-W., Schlafly, E.~F. et al. 2016, \apj, 817, 1, 73
\bibitem[Ibata et al. (1994)]{Ibata1994} Ibata, R.~A., Gilmore, G., Irwin, M.~J., 1994, Nature, 370, 6486, 194
\bibitem[Ibata \& Lewis (1998)]{Ibata1998} Ibata, R., Lewis, G.~F., 1998, \apj, 500, 575
\bibitem[Johnston et al. (1995)]{Johnston1995} Johnston, K.~V., Spergel, D.~N., Hernquist, L., 1995, \apj, 451, 598
\bibitem[Juri{\'{c}} et al. (2008)]{Juric2008} Juri{\'{c}}, M., Ivezi{\'{c}},{\v{Z}}., Brooks, A., et al. 2008, \apj, 673, 864
\bibitem[Kaiser et al. (2010)]{Kaiser2010} Kaiser, N. et al. 2010, Proc. SPIE, 7733
\bibitem[Karachentsev et al. (2004)]{Karachentsev2004} Karachentsev, I.~D., Karachentseva, V.~E., Huchtmeier, W.~K., et al. 2004, \aj, 127, 4, 2031	
\bibitem[Koposov et al. (2010)]{Koposov2010} Koposov, S.~E., Rix, H.~W., Hogg, D.~W., et al. 2010, \apj, 712, 1, 260
\bibitem[Koposov et al. (2012)]{Koposov2012} Koposov, S.~E., Belokurov, V., Evans, N.~E., et al. 2012, \apj, 750, 1, 80
\bibitem[Law et al. (2005)]{Law2005a} Law., D.~R.,Johnston, K.~V.,  Majewski, S.~R., 2005, \apj, 619, 800
\bibitem[Law \& Majewski (2005)] {Law2005b} Law, D.~R., Majewski, S.~R. 2005, \apj, 619, 807
\bibitem[Law \& Majewski (2010)] {Law2010} Law, D.~R., Majewski, S.~R. 2010, \apj, 714, 1, 229
\bibitem[Magnier et al. (2008)]{Magnier2008} Magnier, E.~A., Liu, M., et al. 2008, IAU Symposium, Vol. 248, 553-559
\bibitem[Majewski et al.(2003)]{Majewski2003} Majewski, S.~R., Skrutskie, M.~F., Weinberg, M.~D., \& Ostheimer, J.~C.\ 2003, \apj, 599, 1082
\bibitem[Mart\'{i}nez-Delgardo et al.(2001)]{Martinez2001} Mart\'{i}nez-Delgado, D., Aparicio, A., G\'{o}mez-Flechoso, M. A., et al. 2001, \apjl, 549, 2
\bibitem[Mart\'{i}nez-Delgardo et al.(2004)]{Martinez2004} Mart\'{i}nez-Delgado, D., Aparicio, A., Carrera, Ricardo, et al. 2003, \apj, 601, 1, 242
\bibitem[Mateo et al.(1998)]{Mateo1998} Mateo, M., Olszewski, E.~W., Morrison, H.~L. 1998, ApJL, 508, 1, L55
\bibitem[Navarrete et al.(2017)]{Navarrete2017} Navarrete, C., Belokurov, V., Koposov, E. E., et. al. 2017, \mnras, 467, 2, 1329
\bibitem[Newberg et al.(2002)]{Newberg2002} Newberg, H.~J., Yanny, B., Grebel, E.~K., et al. 2003, ApJL, 596, L191
\bibitem[Newberg et al.(2009)]{Newberg2009} Newberg, H.~J., Yanny, B., Willet, B.~A., 2009, ApJL, 700, 2, L61
\bibitem[Newberg et al. (2010)]{Newberg2010} Newberg, H.~J., Willett, B.~A., Yanny, B., et al. 2010, \apj, 771, 1, 32
\bibitem[Niederste-Ostholt et al. (2010)]{NiedersteOstholt2010} Niederste-Ostholt, M., Belokurov, V., Evans, N.~W., et al. 2010, \apj, 712, 1, 516
\bibitem[Pe{\~{n}}arrubia et al. (2010)] {Penarrubia2010}  Pe{\~{n}}arrubia, J., Belokurov, V., Evans, N.~W., et al. 2010, \mnras, 408, 1, L26
\bibitem[Ruhland et al.(2011)]{Ruhland2011} Ruhland, C., Bell, E.~F., Rix, H.-W., \& Xue, X.-X.\ 2011, \apj, 731, 119 
\bibitem[Sanders \& Binney (2013)]{Sanders2013} Sanders, J.~L., \& Binney, J. 2013, \mnras, 433, 1826
\bibitem[Schlafly et al. (2012)]{Schlafly2012} Schlafly, E.~F., Finkbeiner, D.~F., et al. 2011, \apj, 756, 158 
\bibitem[Sesar et al. (2012)]{Sesar2012} Sesar, B., Cohen, J.~G., Levitan, D., et al. 2012 \apj, 755, 134
\bibitem[Sesar et al. (2013b)]{Sesar2013b} Sesar, B., Ivezi{\'{c}},{\v{Z}}., Scott, J. S., et al.  2013b, \aj, 146, 2, 21
\bibitem[Sesar et al. (2016)]{Sesar2016} Sesar, B., Bovy, J., Bernard, E.~J., et al.  2016, \apj, 809, 1, 59
\bibitem[Sesar et al.(2017a)]{Sesar2017a} Sesar, B., Fouesneau, M., Price-Whelan, A.~M., et al.\ 2017a, \apj, 838, 107
\bibitem[Sesar et al.(2017b)]{Sesar2017b} Sesar, B., Hernitschek, N., Mitrovi{\'c}, S., et al.\ 2017b, \aj, 153, 204 
\bibitem[Sesar et al.(2017c)]{Sesar2017c} Sesar, B., Hernitschek, N., Dierickx, M.~I.~P., Fardal, M.~A., \& Rix, H.-W.\ 2017c, arXiv:1706.10187
\bibitem[Slater et al.(2013)]{Slater2013} Slater, C.~T., Bell, E.~F., Schlafly, E.~F., et al, 2013, \apj, 762, 1, 6
\bibitem[Stubbs et al. (2010)]{Stubbs2010} Stubbs, C.~W., Doherty, P., et al. 2010, Astrophys. J. Suppl. Ser., 191, 376
\bibitem[Tonry et al. (2012)]{Tonry2012} Tonry, C.~W., Stubbs, K. R., et al. 2012, \apj, 750, 99
\bibitem[Vivas et al. (2001)]{Vivas2001} Vivas, A.~K., Zinn, R., Andrews, P., et al. 2001, ApJL, 554, L33
\bibitem[Vivas et al. (2004)]{Vivas2004} Vivas, A.~K., Zinn, R., Abadl, C., et al. 2004, \apj, 127, 2
\bibitem[Vivas et al. (2008)]{Vivas2008} Vivas, A.~K., Jaff\'{e}l, Y.~L., Zinn, R., et al. 2008, \apj, 136, 4
\bibitem[Zinn et al. (2014)]{Zinn2014} Zinn, R., Horowitz, B., Vivas, A.~ K., et al. 2014, \apj, 781, 22

\end{thebibliography}
\end{document}